\documentclass[11pt,letterpaper]{article}
\usepackage[margin=1in]{geometry}
\usepackage{graphicx,amsmath,amsfonts,amssymb,amsthm,bbold,bm,dsfont,cite,xcolor,fullpage,braket,tikz}
\usepackage[labelfont=bf]{caption}
\usepackage{subcaption}
\usepackage[colorlinks,allcolors=red]{hyperref}
\usepackage[capitalize]{cleveref}
\usepackage[title]{appendix}
\usepackage{thmtools} 
\usepackage{diagbox,placeins}
\usepackage{enumitem}
\usepackage{comment}
\usepackage{authblk}
\makeatletter
\renewcommand{\AB@authnote}[1]{{\textsuperscript{\normalfont#1}}}
\renewcommand{\AB@affilnote}[1]{{\textsuperscript{\normalfont#1}}}
\makeatother

\graphicspath{{./figures/}}

\crefformat{equation}{(#2#1#3)}
\Crefformat{equation}{(#2#1#3)}
\crefname{section}{}{\S\S}
\crefformat{section}{#2\S#1#3}
\Crefformat{section}{#2\S#1#3}
\crefname{appendix}{Appendix}{Appendices}
\Crefname{appendix}{Appendix}{Appendices}

\newtheorem{definition}{Definition}

\newtheorem{theorem}{Theorem}
\newtheorem{corollary}[theorem]{Corollary}

\newtheorem{proposition}[theorem]{Proposition}

\newcommand{\FF}{\mathbb{F}}
\newcommand{\GG}{\mathbb{G}}
\newcommand{\dyn}{{\mathrm{dyn}}}
\newcommand{\vect}[1]{{\mathbf{#1}}}
\newcommand{\xv}{{\mathbf{x}}}
\newcommand{\qv}{{\boldsymbol{q}}}
\newcommand{\ev}{{\boldsymbol{e}}}

\newcommand{\etav}{{\boldsymbol{\eta}}}
\newcommand{\muv}{{\boldsymbol\mu}}
\newcommand{\iid}{\mathrm{i.i.d.}}
\newcommand{\reg}{\mathrm{reg}}

\newcommand{\ie}{{i.e.}}

\begin{document}

\title{Regev's Reduction Reaches Reconstruction}

\author[1]{Stephen P.~Jordan}
\author[1,2]{Quynh T.~Nguyen}
\author[3]{Leo Zhou}

\affil[1]{\small{\it{Google Quantum AI, Venice, CA, USA}}}
\affil[2]{\small{\it{Harvard University, Cambridge, MA, USA}}}
\affil[3]{\small{\it{University of California, Los Angeles, CA, USA}}}

\date{November 5, 2026}

\maketitle

\begin{abstract}
Regev's reduction uses a quantum Fourier transform to carry out reductions between various problems related to lattices and codes. We apply it to the problem of sampling from Gibbs distributions of certain classical spin glass Hamiltonians, which can equivalently be thought of as cost functions for average-case instances of an optimization problem called $D$-regular max-$k$-LINSAT. This problem becomes hard at low temperature. Here, we use a quantum decoding algorithm called Belief Propagation with Quantum Messages (BPQM) to solve the decoding problem arising from Regev's reduction. Our main contribution is a proof that the minimum temperature reachable using Regev+BPQM is exactly equal to the reconstruction threshold temperature for noisy channels transmitting signals on a tree network. This equality can be understood via Renes' duality between classical and quantum channels~\cite{R18}, which we use to show that quantum decoding is possible exactly when classical reconstruction on the dual problem is impossible. M{\'e}zard and Montanari~\cite{MM06} identified this reconstruction temperature with a dynamical glass transition predicted by the one-step replica-symmetry breaking (1RSB) cavity equation. It is conjectured that local samplers such as Metropolis Monte Carlo and Glauber dynamics cannot equilibrate efficiently below this reconstruction temperature but may do so above it; our large numerical experiments show close agreement with this prediction across hundreds of different problem parameters. Therefore, our results rigorously show that Regev+BPQM reaches a temperature for which, to our knowledge, there is no matching classical guarantee, and brings Regev+BPQM to the conjectured limit of efficient Gibbs sampling by classical local dynamics.
\end{abstract}

\section{Introduction}

In the early 2000s it was shown by Regev, Aharonov, and Ta-Shma \cite{ATS03,R04,AR05,R09} that the quantum Fourier transform can be used to efficiently realize reductions between various difficult computational problems regarding high dimensional lattices. However, at the time, these reductions did not lead to any quantum algorithms achieving advantage. Recently, a quantum algorithm called Decoded Quantum Interferometry (DQI)~\cite{JSW25} was introduced which in polynomial time finds approximate solutions to a problem called Optimal Polynomial Intersection. No polynomial time classical algorithms are currently known that can find these approximate optima. DQI is closely related to Regev's reduction. In fact, as later shown in \cite{SMR26}, when applied to the same problems using the same decoders, Regev's reduction efficiently achieves the same asymptotic approximate optima as DQI, although DQI outperforms at finite instance sizes. The key new element in \cite{JSW25} to achieving quantum advantage was considering optimization problems with algebraic structure and using decoders that exploited that algebraic structure.

It was also proposed in \cite{JSW25} to consider optimization problems with sparsity structure, such as $D$-regular max-$k$-XORSAT and its larger-alphabet generalization to $D$-regular max-$k$-LINSAT. Via the quantum Fourier transform, these are reduced to problems of decoding Low-Density Parity-Check (LDPC) codes. The low density (\ie, sparsity) of the parity check matrices makes the decoding problem much easier. Known classical algorithms such as belief propagation can efficiently decode LDPC codes of blocklength $m$ even when a (sufficiently small) constant fraction of the bits have been corrupted by bit flips. In contrast, for general dense parity check matrices, no polynomial time classical decoding algorithm is known beyond logarithmically many bit flips. Unfortunately, sparsity also makes the original optimization problem easier. Using classical belief propagation as the decoding algorithm in DQI appears insufficient to achieve quantum advantage on $D$-regular max-$k$-LINSAT over simulated annealing.

In this manuscript, we analyze the performance of Regev's reduction when combined with an intrinsically quantum decoding method called Belief Propagation with Quantum Messages (BPQM) introduced in \cite{R17} and further developed in \cite{PR22,BMP22,MP26}. In \cite{piveteau2026gibbs} this combination was used to prepare Gibbs states of quantum Hamiltonians, including for dilute Ising spin glasses. Here, we apply Regev+BPQM to a classical optimization problem called average-case $D$-regular max-$k$-LINSAT. An instance of $D$-regular max-$k$-LINSAT consists of $n$ variables from $\mathbb{F}_p$, contained in $m$ constraints. For a given assignment $\mathbf{x} \in \mathbb{F}_p^n$ to the variables we can assign a corresponding energy $E(\mathbf{x})$, equal to the number of violated constraints. Given an inverse temperature $\beta\ge 0$ this defines a Gibbs distribution
\begin{equation}
    \label{eq:Gibbs}
    \mathrm{Pr}_{\beta}(\mathbf{x}) = \frac{1}{Z} e^{- \beta E(\mathbf{x})} \quad \textrm{where} \quad Z = \sum_{\mathbf{x} \in \mathbb{F}_p^n} e^{- \beta E(\mathbf{x})}.
\end{equation}
Here, we use Regev's reduction together with BPQM to prepare the state
\begin{equation}
    \label{eq:psi_Gibbs}
    \ket{\psi_\beta} = \sum_{\mathbf{x} \in \mathbb{F}_p^n} \sqrt{\mathrm{Pr}_{\beta}(\mathbf{x})} \ket{\mathbf{x}}.
\end{equation}
That is, we use it to achieve \emph{coherent Gibbs sampling}.
For the random regular ensemble defined in \cref{sec:ensemble}, our algorithm (\cref{cor:finite-sampling}) prepares this state with trace distance $\varepsilon$ tending to zero with high probability as $n\to\infty$, for each fixed $\beta$ strictly below the decoding threshold.

Measuring the state $\ket{\psi_\beta}$ in the computational basis yields classical Gibbs sampling (to within $\varepsilon$ total variation distance) at inverse temperature $\beta$. At low temperature the expected energy $\langle E \rangle_\beta$ of the sampled symbol strings is low, and hence the Gibbs sampler can be used as an optimization algorithm. We remark that coherent Gibbs sampling at inverse temperature $\beta$ is a more powerful primitive than classical Gibbs sampling at inverse temperature $\beta$~\cite{ATS03, Montanaro15, KO22}, which in turn is a more powerful primitive than finding an individual solution at energy $\langle E \rangle_\beta$~\cite{JVV86}.

As our main result, we prove that the BPQM tree-decoding threshold is exactly equal to the reconstruction threshold for a noisy channel on the dual tree network (\cref{thm:main-informal}). The latter is known to correspond to the dynamical glass transition temperature $\beta_\dyn$ in the associated statistical physics model of random max-LINSAT Hamiltonians~\cite{MM06}, beyond which Gibbs sampling is conjectured to be difficult (see e.g., \cite{KT87dynamics,Franz1995,MS06,Alaoui2025sampling}). 
This equality also has a nice information-theoretic interpretation: Renes' channel duality relates the quantum messages used for decoding to classical noisy observations used for reconstructing the root symbol.
In our setting, the quantum messages become perfectly distinguishable precisely when the dual classical observations lose all information about the root.

Our result guarantees coherent Gibbs sampling by Regev+BPQM all fixed $\beta<\beta_\dyn$, reaching the conjectured limit for classical local dynamics~\cite{MS06}. Although it is a folklore belief that Glauber dynamics and other local Markov chains can reach this limit, we are not aware of a matching classical sampling theorem for the general sparse max-LINSAT ensembles considered here. For example, it was only very recently shown that simulated annealing samples up to criticality for the Sherrington-Kirkpatrick model (i.e., random weighted max-$2$-XORSAT on dense graphs)~\cite{hou2026simulated,du2026sampling}. In the more general $k\ge 3$ case, however, existing fast-mixing guarantees for Glauber dynamics only apply at inverse temperatures far smaller than $\beta_\dyn$ of the related dense models \cite{Anari2024,Adhikari2024,Alaoui2025sampling}.
Therefore, our result constitutes the first rigorous proof of a polynomial-time algorithm to produce (coherent) Gibbs samples for all $\beta \in [0, \beta_\dyn)$ for these broad families of spin glass Hamiltonians.

\section{Summary of Results}
\label{sec:problem}

We begin by giving a formal definition of the max-LINSAT problem, which follows \cite{JSW25}:
\begin{definition}
    Let $\mathbb{F}_p$ be a finite field and let $B \in \mathbb{F}_p^{m \times n}$. For each $i = 1,\ldots,m$ let $F_i \subset \mathbb{F}_p$ be an arbitrary subset of $\mathbb{F}_p$ which yields a corresponding constraint $\sum_{j=1}^n B_{ij} x_j \in F_i$. Given $B$ and $\vect{F} = (F_1,\ldots,F_m)$, the \emph{max-LINSAT problem} is to find $\mathbf{x} \in \mathbb{F}_p^n$ satisfying as many as possible of these $m$ constraints.
\end{definition} 

For simplicity, we restrict $p$ to be prime in this work. In the case of $p=2$ and $|F_i| = 1$, one recovers max-XORSAT. An instance of max-LINSAT corresponds to the energy function $E:\mathbb{F}_p^n \to \mathbb{Z}$ given by
\begin{equation}
    \label{eq:EGibbs}
    E(\mathbf{x}) = | \{i: \sum_{j=1}^n B_{ij} x_j \notin F_i\} |.
\end{equation}
This in turn defines a classical Gibbs distribution via \cref{eq:Gibbs} and a coherent Gibbs sampling problem via \cref{eq:psi_Gibbs}. At $p=2$, the energy function \cref{eq:EGibbs} is equivalent to the classical Ising model.

The special case of max-LINSAT where each constraint contains exactly $k$ variables is called max-$k$-LINSAT. That is, in max-$k$-LINSAT each row of the matrix $B$ contains exactly $k$ nonzero coefficients. If each variable is contained in exactly $D$ constraints, \ie, each column of $B$ contains exactly $D$ nonzero coefficients, then the instance is said to be $D$-regular. When $k=2$ and $F_i = \FF_p\setminus\{0\}$, the problem is equivalent to max-$p$-cut or the antiferromagnetic Potts model.

\paragraph{Our Ensemble of Instances.}
To fully specify an instance of max-LINSAT we need to make three choices: which entries of $B$ are nonzero, what values in $\mathbb{F}_p^{\times}$ these nonzero entries take, and what subsets of $\mathbb{F}_p$ we take as our targets $F_1,\ldots,F_m$. We address these in turn.

To generate an instance of $D$-regular max-$k$-LINSAT we start by choosing a matrix $\bar{B}$ uniformly at random from all $\{0,1\}$-valued $m \times n$ matrices with exactly $D$ ones in each column and $k$ ones in each row. The ones represent the locations of nonzero entries in our final matrix $B \in \mathbb{F}_p^{m \times n}$. We can visualize $\bar{B}$ by a corresponding bipartite graph with one part of $m$ vertices that correspond to the constraints in our max-LINSAT instance and a second part of $n$ vertices that correspond to the variables. A constraint vertex $i$ in the first part is connected by an edge to a variable vertex $j$ in the second part if and only if $\bar{B}_{ij} = 1$, \ie, constraint $i$ contains variable $j$. Following the terminology from coding theory, we refer to this as the Tanner graph. 
Given $\bar{B}$, we next construct $B$ by filling in each nonzero entry with a choice from $\mathbb{F}_p^{\times}$, followed by choosing target sets $F_1,\ldots,F_m$.

\begin{definition}[Random regular max-LINSAT]
\label{sec:ensemble}
Fix a prime $p$, integers $3\leq k<D$, a nonempty proper subset $S\subsetneq\mathbb F_p$, and an exchangeable probability law $\Lambda$ on $(\FF_p^\times)^D$. Hold these parameters fixed as $n\to\infty$ along values for which $m=nD/k$ is an integer.
Then a max-LINSAT instance $(B,\vect{F})$ is drawn from the \emph{random regular ensemble $\GG_{\reg}(n,k,D,p,S,\Lambda)$} as follows. First choose $\bar{B}\in \{0,1\}^{m\times n}$ as above; then $B$ is constructed by taking $\bar{B}$ and filling the $D$ nonzero entries in each column one tuple drawn from the law $\Lambda$.
To obtain $\vect{F}=(F_i)_{i=1}^m$, we first choose offsets $\tau_1,\ldots,\tau_m$ independently and uniformly at random from $\FF_p$, and set $F_i=\tau_i+S$.
\end{definition}

Requiring $\Lambda$ to be exchangeable, which means the probability law is invariant under permuting the $D$ variables, is a proof convenience.
Examples of exchangeable $\Lambda$ include:
\begin{enumerate}
    \item $\Lambda_{\rm iid}$: Choose each coefficient independently and uniformly at random from $\FF_{p}^\times$.
    \item $\Lambda_{\rm perm}$: Fix a primitive element $g\in \FF_p^\times$, and choose the coefficients to be a uniformly random permutation of $(1,g,g^2,\ldots, g^{D-1})$.
\end{enumerate}
In particular, the choice $\Lambda_{\rm iid}$ gives the coefficient law in \cite[Theorem~30]{MPRP26}; our \cref{thm:finite-bpqm} extends its argument to any general exchangeable law and conditions the graph to be simple.
Separately, the uniform offsets supply the random shifts required by the coherent-state error bound \cite[Lemma~4]{SMR26}.

\paragraph{Reconstruction and the Main Result.}
The neighborhood of a uniformly chosen vertex in a random regular Tanner graph is a tree at any fixed radius, with probability tending to one as $n\to\infty$. The associated infinite tree has variable degree $D$ and constraint degree $k$. Its free-boundary Gibbs distribution can be generated by a broadcast: choose the root symbol uniformly from $\mathbb F_p$, then at each constraint jointly generate its $k-1$ child symbols from the conditional Gibbs distribution given the parent. Satisfying assignments to that constraint receive weight $1$, and violating assignments receive weight $e^{-\beta}$. This procedure is specified in \cref{sec:reconstruction}.

The reconstruction problem asks an observer who knows the tree, coefficients, and target offsets to infer the root from the symbols at generation $\ell$, even as $\ell\to\infty$. At high temperature, this distant boundary loses all information about the root. At sufficiently low temperature, information may survive. Writing $X_0$ for the root, $X_{\partial T_\ell}$ for the boundary, and $\mathcal T$ for the known tree data, we define the reconstruction threshold in inverse temperature as
\begin{equation}
    \label{eq:summary-reconstruction}
    \beta_\dyn:=\inf\left\{\beta>0:
       \lim_{\ell\to\infty}I(X_0;X_{\partial T_\ell}\mid\mathcal T)>0\right\}.
\end{equation}
The notation $\beta_\dyn$ reflects its relation to the dynamical transition in spin glass theory, as shown in \cite{MM06}. This equivalence is made explicit by examining the distribution of the root's posterior probabilities as the observed boundary moves farther away. If these probabilities approach the uniform distribution, the boundary loses all information about the root; if they retain a nonzero bias, reconstruction remains possible. M\'ezard and Montanari show that the recursive equation tracking the local posterior probabilities is also the $m=1$ one-step replica-symmetry-breaking (1RSB) equation used to locate the dynamical glass transition~\cite[Proposition~1 and Section~7]{MM06}.
A nontrivial solution to the equation describes persistent information on a distant boundary on the tree. On finite random graphs, this translates to a physical picture where the Gibbs measure is decomposed into regions between which local dynamics move slowly.

On the decoding side, Regev's reduction associates the instance matrix $B$ with the LDPC code $C^\perp=\ker B^T$. Its symbols have degree $k$ and its checks have degree $D$, interchanging the two roles in the reconstruction graph. We review Regev's reduction for Gibbs sampling in \cref{sec:regev}. To solve the associated decoding problem,
BPQM, described in \cref{sec:bpqm}, estimates a code symbol by combining quantum messages in a tree neighborhood of the symbol. The performance of BPQM decoding is classically analyzed by the framework of quantum density evolution (\cref{sec:qde}), which tracks the distributions of these messages and determines the limiting root-symbol error. We write $\beta_{\rm BPQM}$ for the inverse temperature at which this error ceases to vanish as the tree depth grows; its precise definition is given in \cref{eq:bpqm-threshold}.
We note that this threshold is optimal for root-symbol decoding on tree Tanner graphs for the pure-state channels considered here, since BPQM minimizes the decoding error among all collective quantum measurements on the tree’s channel outputs \cite{PR22, MPRP26}.

Our main result establishes a precise equality between $\beta_\mathrm{dyn}$ and $\beta_\mathrm{BPQM}$:
\begin{theorem}[Regev+BPQM reaches reconstruction, informal]
\label{thm:main-informal}
Fix any choice of $(k,D,p,S,\Lambda)$ as in \cref{sec:ensemble}.
Then the reconstruction threshold on the associated tree equals the BPQM decoding threshold on the dual tree:
\vspace{-5pt}
\begin{equation} \label{eq:summary-equal}
\beta_{\rm BPQM}(k,D,p,S,\Lambda)=\beta_{\dyn}(k,D,p,S,\Lambda).
\end{equation}
For every $\beta \in [0, \beta_\dyn)$, there is a polynomial-time quantum algorithm based on Regev's reduction with BPQM that
prepares the coherent Gibbs state \eqref{eq:psi_Gibbs}  of an instance drawn from $\GG_\reg(n, k,D,p,S,\Lambda)$, with vanishing trace-distance error with high probability as $n\to\infty$.
\end{theorem}

The main novelty of \cref{thm:main-informal} is the equality \eqref{eq:summary-equal}, which identifies BPQM success exactly with non-reconstruction on the tree.
For the finite-graph sampling conclusion from decoding (\cref{cor:finite-sampling}, \cref{sec:finite-sampling}), our proof is a straightforward adaptation of recent works \cite{MPRP26, SMR26}. In particular, \cite[Theorem~30]{MPRP26} proves vanishing block error for BPQM with erasure recovery on random regular LDPC codes throughout the tree-decoding success region.
Our adapted \cref{thm:finite-bpqm} makes two modest extensions: it allows any exchangeable coefficient law $\Lambda$, with independent tuples at distinct checks, and conditions the Tanner graph to have no parallel edges.
\cite[Lemma~4]{SMR26} supplies the other ingredient: an average fidelity bound that converts decoding error into coherent-state preparation error in Regev's reduction. The independent uniform offsets $\tau_j$ in our ensemble provide the random shifts required by their bound. 

Our proof of the threshold equality \eqref{eq:summary-equal} can be understood through the lens of Renes' channel duality~\cite{R18} (see \cref{sec:main-theorem}). A BPQM message is a channel that maps a code symbol to a quantum state. By \cite{R18}, this quantum message channel is dual to a classical channel that outputs a probability distribution. We then show that each round of BPQM message-combining operation in the dual classical channel is mathematically equivalent to a step in the reconstruction recursion on local posterior distributions. Renes' entropic uncertainty relation then tells us that the quantum decoding has vanishing error (maximal information) in the infinite tree precisely when the classical dual reconstruction problem has vanishing information, which implies that the BPQM decoding threshold is equal to the reconstruction threshold.

\paragraph{Numerical Experiments.} In the spin glass literature it has been conjectured that the Metropolis and Glauber Markov chains have slow mixing below the reconstruction threshold temperature (e.g., \cite{KT87dynamics,MS06}). Hence, one expects that the energy efficiently reachable in practice by simulated annealing and Glauber methods may coincide, at least approximately, with the expected energy in the Gibbs distribution at the tree reconstruction threshold temperature. Prior computational experiments have provided evidence in support of this in several regimes (e.g., \cite{MR04cooling, MS06, Krzakala2007}). In \cref{sec:experiments} we report our own computer experiments with simulated annealing specifically on our ensemble of $D$-regular max-$k$-LINSAT over an alphabet of size $p$ with $\Lambda=\Lambda_{\rm perm}$ and target set $S_r=\{0,...,r-1\}$. We perform our experiments on random instances with $n=6720$ variables, across an exhaustive grid of ensemble parameters for all $3 \leq k < D \leq 8$ with $p$ ranging across all primes up to 13 and $1 \leq r < p$. Among its $525$ settings, density evolution detects finite $\beta_\dyn$ thresholds at $252$, which we subsequently use for simulated annealing experiments. On this region of the four-dimensional parameter space defined by $(k,D,p,r)$ we find that the energy reached by simulated annealing  (normalized to lie between 0 and 1 as a satisfaction fraction) differs from this prediction by an average of $0.58$ percentage points and at most $2.8$ percentage points. This holds across short anneals ($1.44 \times 10^5$ sweeps) and long anneals ($1.44 \times 10^7$ sweeps). Thus, although such experiments cannot definitively prove or disprove the claim, we do find that this theory is remarkably predictive of practical performance in our simulated annealing experiments.

\medskip
\noindent \textbf{Outlook.} Our work leaves open several future directions:
\begin{itemize}
    \item Can one rigorously prove that classical Glauber dynamics sample in polynomial time at all inverse temperatures below the reconstruction threshold ($\beta<\beta_\dyn$) for all max-LINSAT ensembles? This would give a matching classical polynomial-time sampling guarantee below $\beta_\dyn$.
    \item Our quantum algorithm prepares a coherent Gibbs state, which is a more powerful primitive than classical Gibbs samples. In particular, coherent Gibbs sampling has led to several quadratic quantum speedups in optimization and estimating partition functions~\cite{Montanaro15, SBBK08, WCNA09, KO22, AHNTW22}. Our algorithm might lead to similar speedups.
    \item Can one Gibbs sample beyond the reconstruction threshold for these random max-LINSAT problems? The recent work~\cite{ZSSJ26} gave such classical and quantum algorithms in certain regimes of large $(k,D)$ for the special case of max-XORSAT, but those algorithms underperform compared to Regev+BPQM for the small $(k,D)$ regime we consider here.
	A quantum algorithm that Gibbs samples at $\beta>\beta_\dyn$ over wider regimes of $(k,D,p,S,\Lambda)$ could lead to a quantum advantage in sampling and optimization for unstructured spin glass problems.
\end{itemize}

\noindent \textbf{Concurrent Work.} While we were preparing this manuscript,
we learned that our main result, in particular the correspondence between Regev+BPQM and the reconstruction threshold, was independently and simultaneously obtained by the Global Technology Applied Research team at JPMorganChase~\cite{augustino2026tree}.

\section{Regev's Reduction}
\label{sec:regev}

We now briefly review Regev's reduction from a point of view suited to present purposes. 

A matrix $B \in \mathbb{F}_p^{m \times n}$ defines two codes:
\begin{equation}
    \label{eq:codes}
    C = \{B \mathbf{x} : \mathbf{x} \in \mathbb{F}_p^n\} \quad \mathrm{and} \quad C^{\perp} = \{ \mathbf{d} \in \mathbb{F}_p^m : B^T \mathbf{d} = \mathbf{0} \}.
\end{equation}
Their normalized code states are
\begin{equation}
    \ket{C} = \frac{1}{\sqrt{|C|}}\sum_{\mathbf{c}\in C}\ket{\mathbf{c}},
    \qquad
    \ket{C^\perp} = \frac{1}{\sqrt{|C^\perp|}}\sum_{\mathbf{d}\in C^\perp}\ket{\mathbf{d}}.
\end{equation}
It is easy to verify that $\mathrm{QFT}\ket{C}=\ket{C^\perp}$ and $\mathrm{QFT}^{\dagger}\ket{C}=\ket{C^\perp}$, where $\mathrm{QFT}$ denotes the unitary Fourier transform over $\mathbb{F}_p^m$.

Now suppose that, for an envelope function $P:\mathbb{F}_p^m\to\mathbb{C}$, we wish to construct the state
\begin{equation}
    \label{eq:psi}
    \ket{\psi} = \frac{1}{\sqrt{Z_C(P)}} \sum_{\mathbf{c} \in C} P(\mathbf{c}) \ket{\mathbf{c}},
\end{equation}
where
\begin{equation}
    Z_C(P)=\sum_{\mathbf{c}\in C}|P(\mathbf{c})|^2.
\end{equation}
The Fourier convolution theorem gives
\begin{equation}
    \label{eq:tildepsi}
    \ket{\widetilde{\psi}} := \mathrm{QFT}\ket{\psi}
    = \frac{1}{|C^\perp|\sqrt{Z_C(P)}} \sum_{\mathbf{d} \in C^\perp} \sum_{\mathbf{e} \in \mathbb{F}_p^m} \widetilde{P}(\mathbf{e})\ket{\mathbf{d}+\mathbf{e}},
\end{equation}
where
\begin{equation}
\label{eq:QFT}
    \widetilde{P}(\mathbf{e})
    = \frac{1}{p^{m/2}}\sum_{\mathbf{y}\in\mathbb{F}_p^m}
      \exp\!\left(\frac{2\pi i\langle \mathbf{e},\mathbf{y}\rangle}{p}\right)P(\mathbf{y})
\end{equation}
is the unitary Fourier transform of $P$. The ideal form of Regev's reduction proceeds as follows; \cref{sec:finite-sampling} gives the agreement test and error bound for an approximate decoder.
\begin{enumerate}
    \item Prepare
    \begin{equation}
    \left( \frac{1}{\sqrt{|C^\perp|}} \sum_{\mathbf{d} \in C^\perp} \ket{\mathbf{d}}\right)
    \otimes
    \left( \frac{1}{\|P\|_2}\sum_{\mathbf{e} \in \mathbb{F}_p^m} \widetilde{P}(\mathbf{e}) \ket{\mathbf{e}} \right),
    \end{equation}
    where $\|P\|_2^2=\sum_{\mathbf{y}\in\mathbb{F}_p^m}|P(\mathbf{y})|^2$. The first register is efficiently preparable for any linear code $C^\perp$, while the second is normalized by Parseval's identity. For $P=P_\beta$, its amplitudes factorize over the $m$ coordinates, so it is a product state and hence easy to prepare.
    \item Reversibly add the content of the first register to the content of the second register.
    \begin{equation}
    \to \frac{1}{\sqrt{|C^\perp|}\,\|P\|_2}
    \sum_{\mathbf{d} \in C^\perp} \ket{\mathbf{d}}
    \sum_{\mathbf{e} \in \mathbb{F}_p^m} \widetilde{P}(\mathbf{e}) \ket{\mathbf{d} + \mathbf{e}}.
    \end{equation}

    \item \label{it:decoding-step}
    From the content $\mathbf{d} + \mathbf{e}$ of the second register, coherently infer the content $\mathbf{d}$ of the first register. This is the quantum decoding problem: given the state
    \begin{equation} \label{eq:corrupted-codeword}
        \frac{1}{\|P\|_2}\sum_{\mathbf{e} \in \mathbb{F}_p^m}
        \widetilde{P}(\mathbf{e}) \ket{\mathbf{d} + \mathbf{e}},
    \end{equation}
    which we interpret as a codeword $\mathbf{d} \in C^\perp$ corrupted with a coherent superposition of errors $\mathbf{e}$, infer $\mathbf{d}$. The aim is to erase the first register and uncompute the decoder's work registers. The normalized target of this step is
    \begin{equation}
        \to \ket{\mathbf{0}}\otimes
        \frac{1}{|C^\perp|\sqrt{Z_C(P)}}
        \sum_{\mathbf{d} \in C^\perp}
        \sum_{\mathbf{e} \in \mathbb{F}_p^m}
        \widetilde{P}(\mathbf{e}) \ket{\mathbf{d} + \mathbf{e}}
        = \ket{\mathbf{0}}\otimes\ket{\widetilde{\psi}}.
    \end{equation}

    \item Discard the $\ket{\mathbf{0}}$ register and perform an inverse quantum Fourier transform. By \cref{eq:psi} and \cref{eq:tildepsi} this yields the desired final state.
    \begin{equation}
        \to \frac{1}{\sqrt{Z_C(P)}} \sum_{\mathbf{c} \in C} P(\mathbf{c}) \ket{\mathbf{c}}
        = \ket{\psi}.
    \end{equation}
\end{enumerate}

We next observe how this applies to coherent Gibbs sampling. Every $\mathbf c\in C$ has exactly $|\ker B|$ preimages, all with the same energy. Consequently, the normalized code state corresponding to \cref{eq:psi_Gibbs} is
\begin{equation}
    \label{eq:Cbeta}
    \ket{C_\beta}
       =\frac{1}{\sqrt{|\ker B|}}
          \sum_{\mathbf x\in\mathbb F_p^n}
             \sqrt{\mathrm{Pr}_\beta(\mathbf x)}\ket{B\mathbf x}.
\end{equation}
Gaussian elimination gives a particular preimage of each codeword and a basis for $\ker B$. Preparing a uniform superposition over that kernel implements the isometry
\begin{equation}
    \label{eq:fiber-isometry}
    \ket{\mathbf c}\ket{\mathbf0}
       \longmapsto
    \ket{\mathbf0}\,
       \frac{1}{\sqrt{|\ker B|}}
          \sum_{\mathbf x:\,B\mathbf x=\mathbf c}\ket{\mathbf x},
    \qquad \mathbf c\in C.
\end{equation}
A reversible linear-algebra circuit extends this map to the full registers. Applying it to \cref{eq:Cbeta} gives $\ket{\mathbf0}\ket{\psi_\beta}$. Thus no full-column-rank assumption is needed; when $\ker B=\{\mathbf0\}$, the map reduces to the inverse encoding $\mathbf c\mapsto\mathbf x$.

By \cref{eq:Gibbs} and \cref{eq:codes} we can rewrite \cref{eq:Cbeta} as
\begin{equation}
    \ket{C_\beta} = \sum_{\mathbf{c} \in C} P_\beta(\mathbf{c}) \ket{\mathbf{c}},
\end{equation}
where, for $\mathbf{y}\in\mathbb{F}_p^m$ and $a\in\mathbb{F}_p$,
\begin{align}
    P_\beta(\mathbf{y})
    &= \sqrt{\frac{|\ker B|}{Z}} \prod_{j=1}^m P_\beta^{(j)}(y_j), \label{eq:P1}\\
    P_\beta^{(j)}(a)
    &= \begin{cases}
        1, & a\in F_j,\\
        e^{-\beta/2}, & a\notin F_j.
    \end{cases} \label{eq:P2}
\end{align}
Counting the $|\ker B|$ preimages of each codeword gives $Z_C(P_\beta)=1$. Thus \cref{eq:tildepsi} specializes to
\begin{equation}
    \label{eq:noisy-dual-codeword}
    \ket{\widetilde{C}_\beta} := \mathrm{QFT}\ket{C_\beta}
    = \frac{1}{|C^\perp|}
      \sum_{\mathbf{d}\in C^\perp}
      \sum_{\mathbf{e}\in \mathbb{F}_p^m}
      \widetilde{P}_\beta(\mathbf{e})\ket{\mathbf{d}+\mathbf{e}},
\end{equation}
where $\widetilde{P}_\beta$ is the unitary Fourier transform of $P_\beta$. Since \cref{eq:P1} factorizes,  applying \eqref{eq:QFT} gets
\begin{equation}
    \widetilde{P}_\beta(\mathbf{e}) = \sqrt{\frac{|\ker B|}{Z}} \prod_{j=1}^m \widetilde{P}_\beta^{(j)}(e_j),
\end{equation}
where the individual components \cref{eq:P2} are Fourier transformed to yield
\begin{equation}
    \label{eq:tildeP}
    \widetilde{P}_\beta^{(j)}(e_j)
    = \frac{1}{\sqrt{p}}
      \begin{cases}
        |F_j|+\bigl(p-|F_j|\bigr)e^{-\beta/2}, & e_j=0,\\[2mm]
        \bigl(1-e^{-\beta/2}\bigr)
        \displaystyle\sum_{a\in F_j}
        \exp\!\left(\frac{2\pi i e_j a}{p}\right), & e_j\neq 0.
      \end{cases}
\end{equation}
The amplitudes in \cref{eq:tildeP} define the error model for the quantum decoding problem that we must solve, where the $j$-th codeword symbol is sent through the noise channel 
\[
\textstyle
\ket{d_j}\to \frac{1}{\|P_\beta^{(j)}\|_2}\sum_e \widetilde{P}_\beta^{(j)}(e) \ket{d_j+e}.
\]
In the next section we recount a method for solving this problem called BPQM that was originally developed with quantum communication in mind. Here we find a new application for it in the context of quantum algorithms.

\section{Belief Propagation with Quantum Messages}
\label{sec:bpqm}

We solve the quantum decoding problem in step~\ref{it:decoding-step} of Regev's reduction using Belief Propagation with Quantum Messages (BPQM)~\cite{R17}, in particular its extension to prime alphabets in~\cite{MP26}. The quantum decoding problem is to infer the codeword $\mathbf{d} \in C^\perp$ from the quantum state in \cref{eq:corrupted-codeword}
\begin{align}
    \sum_{\mathbf{e} \in \mathbb{F}_p^m} \widetilde{P}(\mathbf{e})\ket{\mathbf{d}+\mathbf{e}}.
\end{align}

BPQM recovers $\mathbf{d}$ by estimating each code symbol. It estimates a code symbol $d$ by combining messages along a tree rooted at that symbol\cite{R17,PR22}. Each message consists of a quantum system and a classical description of its possible states. There are two combining operations below. Given two input messages, a combining operation produces a single output message. Repeatedly applying the combining operations along the tree produces a single message at the root. A final measurement of the root message estimates the root symbol. Below we recall how BPQM works on a tree; in \cref{sec:finite-sampling} we show how to extend it to our code ensemble.

\paragraph{Messages.}
We use the $p$-ary pure-state channel formulation of \cite[Sec.~II-C, Eq.~(1)]{MP26}, with normalized eigenvalues $\mu_a=\lambda_a/p$ and a fixed change from their Fourier basis to the computational basis. Let $\omega=e^{2\pi i/p}$. A message about a symbol $d\in\mathbb{F}_p$ has a classical component specified by a probability vector $\boldsymbol\mu=(\mu_a)_{a\in\mathbb{F}_p}$ and a quantum component in the state
\begin{equation}
    \label{eq:bpqm-message}
    \ket{\phi_d(\boldsymbol\mu)}
    =\sum_{a\in\mathbb{F}_p}\sqrt{\mu_a}\,\omega^{-ad}\ket{a}.
\end{equation}
The Gram matrix of these states has Fourier eigenvalues $p\mu_a$ \cite[Lemmas~5--7]{MP26}. The normalized eigenvalues $\boldsymbol\mu$ specify the overlaps among the possible states; for example, uniform $\boldsymbol\mu$ gives mutually orthogonal states and perfect distinguishability.

We now show how the quantum message \eqref{eq:bpqm-message} arises in Regev's reduction. For fixed $\mathbf d$, the received state in Regev's reduction~\eqref{eq:noisy-dual-codeword} is a product of the normalized single-symbol received states
\begin{equation}
    \label{eq:regev-symbol-state}
    \ket{v_{d_j}^{(j)}}
    :=\sum_{e\in\mathbb F_p} \frac{\widetilde P_\beta^{(j)}(e)}{\|P_\beta^{(j)}\|_2}\ket{d_j+e}.
\end{equation}
Define the probability vector $\boldsymbol\mu_j$ by normalizing the squared amplitudes of $P_\beta^{(j)}$. Using \cref{eq:P2}, we obtain
\begin{equation}
    \label{eq:bpqm-initial}
    \mu_{j,a}
    =\frac{|P_\beta^{(j)}(a)|^2}{\sum_{b\in\mathbb{F}_p}|P_\beta^{(j)}(b)|^2}
    =\frac{1}{|F_j|+(p-|F_j|)e^{-\beta}}
    \begin{cases}
        1,&a\in F_j,\\
        e^{-\beta},&a\notin F_j.
    \end{cases}
\end{equation}
Applying the inverse Fourier transform to the received state then gives the message state in \cref{eq:bpqm-message} with $d=d_j$ and $\boldsymbol\mu=\boldsymbol\mu_j$:
\[
    \mathrm{QFT}^{\dagger}\ket{v_{d_j}^{(j)}}
    =\sum_{a\in\mathbb F_p}\sqrt{\mu_{j,a}}\,\omega^{-ad_j}\ket a
    =\ket{\phi_{d_j}(\boldsymbol\mu_j)}.
\]

\paragraph{Equality rule.}
Suppose two incoming messages concern the same unknown symbol $d$. The goal of the equality rule\footnote{Also referred to as `variable-node combining' in the literature.} is to compress their two quantum systems into one qudit without losing any information about $d$. This is possible because, conditioned on the classical messages $\boldsymbol\mu$ and $\boldsymbol\nu$, the $p$ possible joint pure states $\{\ket{\phi_d(\boldsymbol\mu)} \otimes \ket{\phi_d(\boldsymbol\nu)} \}_{d \in \mathbb{F}_p}$ span a subspace of dimension at most $p$ inside the $p^2$-dimensional two-qudit space. A unitary can therefore put all dependence on $d$ into one qudit while leaving the other in a fixed state.

To see the compression explicitly, expand the product state and group terms by $a=b+c$:
\[
\begin{aligned}
    \ket{\phi_d(\boldsymbol\mu)}\otimes\ket{\phi_d(\boldsymbol\nu)}
    &=\sum_{b,c\in\mathbb F_p}\sqrt{\mu_b\nu_c}\,
      \omega^{-(b+c)d}\ket{b,c}
    &=\sum_{a\in\mathbb F_p}\omega^{-ad}
      \underbrace{\left(\sum_{b\in\mathbb F_p}
        \sqrt{\mu_b\nu_{a-b}}\ket{b,a-b}\right)}_{\displaystyle\ket{g_a}}.
\end{aligned}
\]
The vectors $\ket{g_a}$ are independent of $d$ and have disjoint computational-basis supports for distinct $a$, so they are mutually orthogonal. Their squared norms give the circular-convolution update of~\cite[Lemma~13]{MP26}, expressed in normalized eigenvalues:
\begin{equation}
    \label{eq:bpqm-equality}
    (\boldsymbol\mu*\boldsymbol\nu)_a
    =\langle g_a|g_a\rangle
    =\sum_{b\in\mathbb{F}_p}\mu_b\nu_{a-b}.
\end{equation}
We can therefore choose a unitary, depending only on $\boldsymbol\mu$ and $\boldsymbol\nu$, that maps each normalized vector to $\ket a\otimes\ket0$. Equivalently,
\[
    V_{\mathrm{eq}}(\boldsymbol\mu,\boldsymbol\nu)\ket{g_a}
    =\sqrt{(\boldsymbol\mu*\boldsymbol\nu)_a}\,\ket a\otimes\ket0.
\]
Applying this unitary to the expanded input state gives
\begin{equation}
    \label{eq:bpqm-equality-unitary}
    \begin{aligned}
    V_{\mathrm{eq}}(\boldsymbol\mu,\boldsymbol\nu)
    \bigl(\ket{\phi_d(\boldsymbol\mu)}\otimes\ket{\phi_d(\boldsymbol\nu)}\bigr)
    &=\left(\sum_{a\in\mathbb F_p}
      \sqrt{(\boldsymbol\mu*\boldsymbol\nu)_a}\,\omega^{-ad}\ket a\right)\otimes\ket0
    &=\ket{\phi_d(\boldsymbol\mu*\boldsymbol\nu)}\otimes\ket{0}.
    \end{aligned}
\end{equation}
This implements the bit-node compression specified in \cite[Sec.~III-A, Eq.~(3), and Remark~14]{MP26} in our message basis. The second output is fixed and can be discarded without loss of information about $d$. The classical update is deterministic.

\paragraph{Parity rule.}
Suppose two incoming messages concern unknown symbols $d_1$ and $d_2$, respectively. The goal of the parity rule is to combine them into a message about their sum $d=d_1+d_2$, which is their parity when $p=2$. The two input symbols can vary independently, so their joint states may span the full two-qudit space. We will obtain a message consisting of one qudit together with a classical measurement outcome that specifies its state description.

The idea is to rearrange the input amplitudes so that the phase on the first qudit depends only on $d_1+d_2$. Apply modular subtraction, $V_{\mathrm{par}}\ket{a,b}=\ket{a,b-a}$, which is a CNOT for $p=2$. Writing $s=b-a$, the resulting state is
\begin{equation}
    V_{\mathrm{par}}
    \bigl(\ket{\phi_{d_1}(\boldsymbol\mu)}\otimes\ket{\phi_{d_2}(\boldsymbol\nu)}\bigr)
    =\sum_{s\in\mathbb F_p}\omega^{-sd_2}
      \left(\sum_{a\in\mathbb F_p}
        \sqrt{\mu_a\nu_{a+s}}\,\omega^{-a(d_1+d_2)}\ket a\right)\ket s.
\end{equation}
Measuring the second qudit now selects a value of $s$. For each fixed input pair $(d_1,d_2)$, the factor $\omega^{-sd_2}$ is then only a global phase, and the first qudit has exactly the form of a message about $d_1+d_2$. The probability of outcome $s$ and the normalized classical description of this message are
\begin{equation}
    \label{eq:bpqm-parity}
    \pi_s=\sum_{a\in\mathbb{F}_p}\mu_a\nu_{a+s},
    \qquad
    \mu_a^{(s)}=\frac{\mu_a\nu_{a+s}}{\pi_s}
    \quad(\pi_s>0).
\end{equation}
These formulas are the sum-convention form of the check-node calculation in \cite[Lemma~12 and Appendix~E-A]{MP26}.\footnote{Their Definition~10 uses the difference of the two input symbols. Negating the second input label, with the corresponding relabeling of Fourier and outcome indices, gives our sum convention.} Thus the outgoing quantum state is $\ket{\phi_{d_1+d_2}(\boldsymbol\mu^{(s)})}$. Although the outcome probabilities are independent of the input symbols, we retain $s$ because it specifies which classical description $\boldsymbol\mu^{(s)}$ to use in subsequent combining operations \cite[Remark~15]{MP26}. The classical update therefore depends on the measurement outcome. An ensemble of pure-state messages accompanied by such a known classical label is called a \emph{heralded mixture} \cite[Definition~24]{MP26}.

\paragraph{BPQM decoding on a tree.} For the code $C^\perp$ in \cref{eq:codes}, the Tanner graph has symbol nodes $d_j$ and check nodes enforcing $\sum_j B_{ji}d_j=0$. Symbols have degree $k$ and checks have degree $D$. Choose a finite tree neighborhood rooted at the symbol to be decoded (non-tree neighborhoods will be considered in Section~\ref{sec:finite-sampling}). The goal is to estimate the root symbol by combining the quantum information available throughout a tree. Each symbol $d_j$ has its own received quantum system, \ie, a message $\ket{\phi_{d_j}(\boldsymbol\mu_j)}$ as above. Starting at the leaf symbols, pass messages toward the root. A leaf sends its own message to its parent check. Internal nodes combine the messages received from their children, using the parity rule at checks and the equality rule at symbols, as follows.

At a check node $i$ whose parent is symbol $d_j$, the child messages concern the other symbols in that check. The check equation expresses the parent symbol as
\[
    d_j=\sum_{\substack{h\ne j\\B_{hi}\ne0}}
        \bigl(-B_{ji}^{-1}B_{hi}\bigr)d_h.
\]
To apply the parity rule repeatedly to obtain one message about $d_j$, we first relabel each child's message to describe its term in this sum: For $\alpha\in\mathbb F_p^\times$, a relabeling $d_h\mapsto\alpha d_h$ is implemented by $\ket a\mapsto\ket{\alpha^{-1}a}$ and $\mu_a\mapsto\mu_{\alpha a}$.

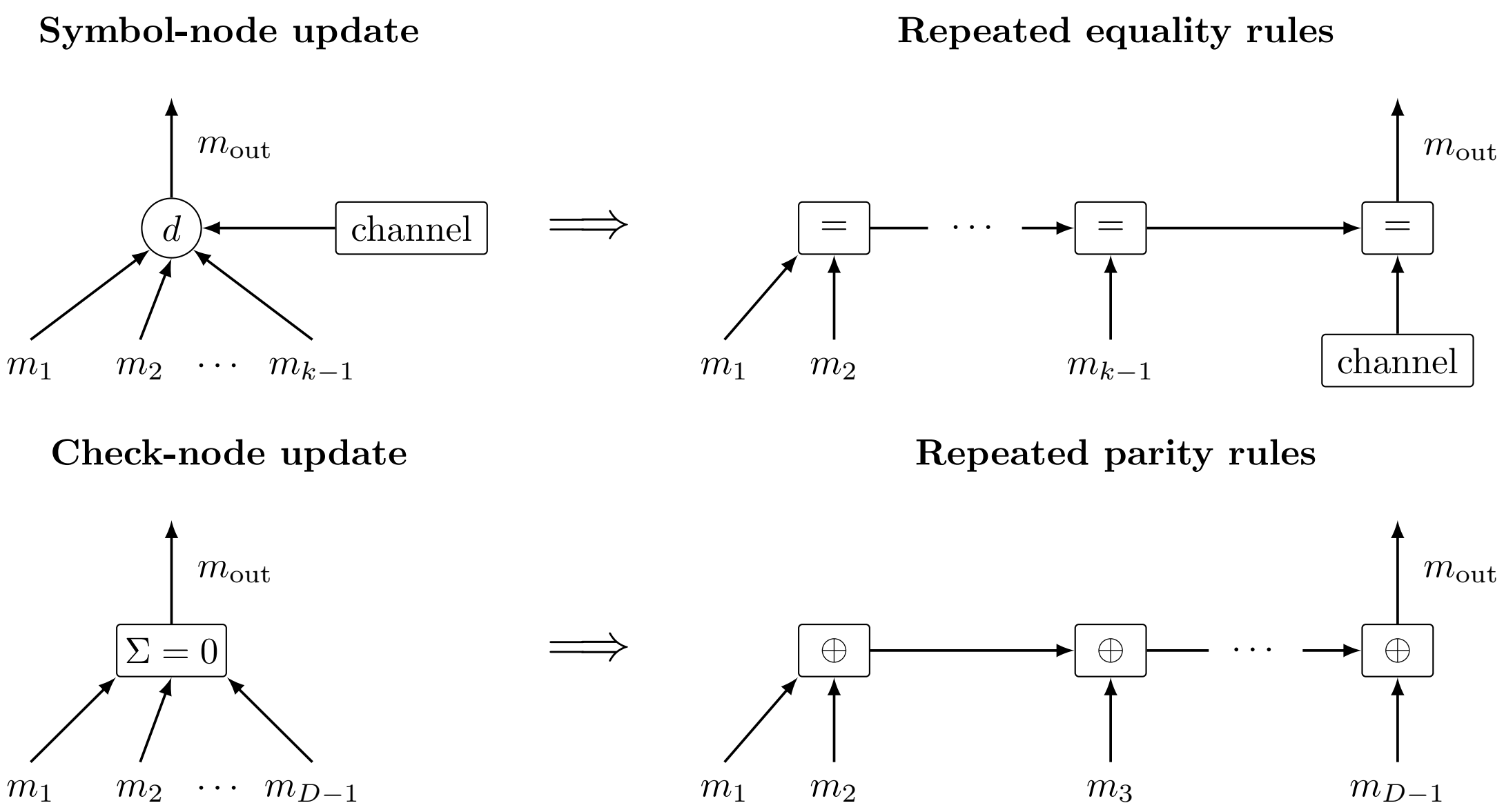
\begin{figure}[t]
    \centering
    \begin{tikzpicture}[
        x=1cm,y=1cm,>=latex,font=\small,
        msg/.style={->,line width=0.7pt},
        link/.style={line width=0.7pt},
        combine/.style={draw,rounded corners=1pt,minimum width=0.68cm,
            minimum height=0.50cm,inner sep=1pt},
        channel/.style={draw,rounded corners=1pt,minimum width=1.45cm,
            minimum height=0.50cm,inner sep=2pt},
        heading/.style={font=\small\bfseries},
        inputlabel/.style={anchor=north,inner sep=2pt}
    ]
        \path[use as bounding box] (0,-4.40) rectangle (15.65,3.45);

        \node[heading] at (2.65,3.12) {Symbol-node update};
        \node[heading] at (11.15,3.12) {Repeated equality rules};
        \node[draw,circle,minimum size=0.57cm,inner sep=1pt]
            (v) at (2.10,1.25) {$d$};
        \draw[msg] (v.north) -- (2.10,2.50)
            node[midway,right=3pt] {$m_{\mathrm{out}}$};
        \draw[msg] (0.75,0.18) -- (v.south west);
        \draw[msg] (1.80,0.18) -- (v.south);
        \draw[msg] (3.45,0.18) -- (v.south east);
        \node[inputlabel] at (0.75,0.08) {$m_1$};
        \node[inputlabel] at (1.80,0.08) {$m_2$};
        \node[inputlabel] at (2.57,0.08) {$\cdots$};
        \node[inputlabel] at (3.45,0.08) {$m_{k-1}$};
        \node[channel] (vchannel) at (4.40,1.25) {channel};
        \draw[msg] (vchannel.west) -- (v.east);

        \node[font=\Large] at (6.10,1.25) {$\Longrightarrow$};

        \node[combine] (e1) at (8.45,1.25) {$=$};
        \node[combine] (e2) at (11.10,1.25) {$=$};
        \node[combine] (e3) at (13.85,1.25) {$=$};
        \draw[msg] (7.40,0.18) -- (e1.south west);
        \draw[msg] (8.45,0.18) -- (e1.south);
        \node[inputlabel] at (7.40,0.08) {$m_1$};
        \node[inputlabel] at (8.45,0.08) {$m_2$};
        \draw[link] (e1.east) -- (9.35,1.25);
        \node at (9.80,1.25) {$\cdots$};
        \draw[msg] (10.25,1.25) -- (e2.west);
        \draw[msg] (11.10,0.18) -- (e2.south);
        \node[inputlabel] at (11.10,0.08) {$m_{k-1}$};
        \draw[msg] (e2.east) -- (e3.west);
        \node[channel] (echannel) at (13.85,-0.02) {channel};
        \draw[msg] (echannel.north) -- (e3.south);
        \draw[msg] (e3.north) -- (13.85,2.50)
            node[midway,right=3pt] {$m_{\mathrm{out}}$};

        \begin{scope}[yshift=-4.05cm]
            \node[heading] at (2.65,3.12) {Check-node update};
            \node[heading] at (11.15,3.12) {Repeated parity rules};
            \node[combine,minimum width=1.05cm]
                (c) at (2.10,1.25) {$\Sigma=0$};
            \draw[msg] (c.north) -- (2.10,2.50)
                node[midway,right=3pt] {$m_{\mathrm{out}}$};
            \draw[msg] (0.75,0.18) -- (c.south west);
            \draw[msg] (1.80,0.18) -- (c.south);
            \draw[msg] (3.45,0.18) -- (c.south east);
            \node[inputlabel] at (0.75,0.08) {$m_1$};
            \node[inputlabel] at (1.80,0.08) {$m_2$};
            \node[inputlabel] at (2.57,0.08) {$\cdots$};
            \node[inputlabel] at (3.45,0.08) {$m_{D-1}$};

            \node[font=\Large] at (6.10,1.25) {$\Longrightarrow$};

            \node[combine] (p1) at (8.45,1.25) {$\oplus$};
            \node[combine] (p2) at (11.10,1.25) {$\oplus$};
            \node[combine] (p3) at (13.85,1.25) {$\oplus$};
            \draw[msg] (7.40,0.18) -- (p1.south west);
            \draw[msg] (8.45,0.18) -- (p1.south);
            \node[inputlabel] at (7.40,0.08) {$m_1$};
            \node[inputlabel] at (8.45,0.08) {$m_2$};
            \draw[msg] (p1.east) -- (p2.west);
            \draw[msg] (11.10,0.18) -- (p2.south);
            \node[inputlabel] at (11.10,0.08) {$m_3$};
            \draw[link] (p2.east) -- (12.04,1.25);
            \node at (12.49,1.25) {$\cdots$};
            \draw[msg] (12.94,1.25) -- (p3.west);
            \draw[msg] (13.85,0.18) -- (p3.south);
            \node[inputlabel] at (13.85,0.08) {$m_{D-1}$};
            \draw[msg] (p3.north) -- (13.85,2.50)
                node[midway,right=3pt] {$m_{\mathrm{out}}$};
        \end{scope}
    \end{tikzpicture}
    \caption{BPQM updates toward the root. (Top) At a symbol node, equality rules combine $k-1$ child-check messages with the symbol's own noise channel output $\ket{\phi_d(\muv)}$. (Bottom) At a check node, parity rules combine $D-1$ child-symbol messages into a message about the parent symbol, with coefficient relabelings understood. Each arrow carries a quantum system and its classical description. The right panels show successive pairwise applications of the rules.}
    \label{fig:bpqm-splitting}
\end{figure}

At a symbol node $d_j$, all messages from its child checks concern the same symbol. Combine them with the symbol's own received message by repeated applications of the equality rule. A non-root internal symbol combines its $k-1$ child-check messages and sends the result to its parent check; the root combines all $k$ check messages with its own received message. Thus the computation proceeds from the leaves to the root, with each outgoing message summarizing the information in the subtree below it. Nodes with several children simply apply the pairwise rules successively, as illustrated in Figure~\ref{fig:bpqm-splitting} (see also \cite[Sec.~3.1.1]{PR22}).

At the root, the procedure leaves a single qudit in the state $\ket{\phi_{d_{\mathrm{root}}}(\boldsymbol\mu_{\mathrm{root}})}$, together with the classical message $\boldsymbol\mu_{\mathrm{root}}$. The probability vector $\boldsymbol\mu_{\mathrm{root}}$ is known from the classical outcome record and describes the possible states of this final qudit. To estimate $d_{\mathrm{root}}$, apply the pretty good measurement, which achieves error probability \cite[Sec.~II-E and Lemma~9]{MP26}
\begin{equation}
    \label{eq:bpqm-error}
    P_{\mathrm{err}}(\boldsymbol\mu_{\mathrm{root}})
    =1-\frac{1}{p}\left(\sum_{a\in\mathbb{F}_p}\sqrt{\mu_{\mathrm{root},a}}\right)^2.
\end{equation}
Averaging over the classical outcomes, the minimum error for this local decoding problem is therefore 
\begin{equation}
    P_{\mathrm{err}}= \mathbb{E}[P_{\mathrm{err}}(\boldsymbol\mu_{\mathrm{root}})].
\end{equation}
The entire procedure is optimal for decoding the root symbol when the assignment is chosen uniformly from those satisfying the tree's checks \cite[Theorem~19]{MPRP26}.

There are a few more ingredients to turn the above tree BPQM decoder into a global codeword decoder. First, for usage in Regev's reduction, we must keep the outcome registers in parity rules unmeasured and perform the classical updates of $\boldsymbol{\mu}$ and conditional gates coherently. This coherent implementation follows \cite[Sec.~5.2]{PR22} and \cite{MPRP26}.\footnote{These references have worked out the coherent finite-precision arithmetic for binary alphabet, but the coherent arithmetic there extends directly to any fixed alphabet.} Retaining these registers and the other work systems allows the combining circuit to be reversed after extracting the symbol estimate. Second, decoding symbols in overlapping tree neighborhoods requires an analysis of the resulting disturbance of the quantum state. Third, the Tanner graph is not a tree, so we must analyze the effect of cycles on the decoding performance. All these subtleties were recently addressed in~\cite{MPRP26}; we will describe this and apply it to Gibbs sampling in~\cref{sec:finite-sampling}.

Finally, we need a classical algorithm to predict the decoding performance and threshold of BPQM. The random outcomes make $\boldsymbol\mu$ a random probability vector. This can be achieved by recursively tracking the classical laws determined by \cref{eq:bpqm-equality,eq:bpqm-parity}. This is called quantum density evolution~\cite{BMP22,MP26,MPRP26}, which we describe in the next section.

\section{Quantum Density Evolution}
\label{sec:qde}
Quantum density evolution is a classical algorithm that applies the equality and parity update rules of \cref{sec:bpqm} to track the distributions of the classical parts of the BPQM messages toward the root \cite{BMP22,MP26,MPRP26}. Here we record the resulting recursion and fix notation for its comparison with tree reconstruction in \cref{sec:main-theorem}.

\begin{figure}[htbp]
\centering
\begin{tikzpicture}[
    x=1cm,y=0.9cm,>=latex,font=\small,
    msg/.style={->,line width=0.65pt},
    symbol/.style={draw,circle,minimum size=0.60cm,inner sep=1pt},
    check/.style={draw,rectangle,minimum size=0.50cm,inner sep=1pt},
    childcheck/.style={draw,rectangle,minimum size=0.20cm,inner sep=0pt}
]
    \node[font=\footnotesize] at (0,4.20) {toward the root};
    \node[symbol] (parent) at (0,3.60) {$d$};
    \node[check] (check) at (0,2.35) {$\Sigma$};
    \node[symbol] (left) at (-1.55,0.85) {$\phantom{~~}d_1\phantom{~~}$};
    \node[symbol] (right) at (1.55,0.85) {$d_{D-1}$};
    \node at (0,0.85) {$\cdots$};

    \draw[msg] (check) -- (parent)
        node[midway,right=3pt] {$\boldsymbol q^{(\ell+1)}$};
    \draw[msg] (left) -- (check)
        node[midway,left=2pt] {$\boldsymbol v_1^{(\ell)}$};
    \draw[msg] (right) -- (check)
        node[midway,right=2pt] {$\boldsymbol v_{D-1}^{(\ell)}$};
    \draw[msg] (-2.85,0.85) -- (left)
        node[midway,above=2pt] {$\boldsymbol\mu_1$};
    \draw[msg] (2.85,0.85) -- (right)
        node[midway,above=2pt] {$\boldsymbol\mu_{D-1}$};

    \node[childcheck] (ll) at (-2.15,-0.55) {};
    \node[childcheck] (lr) at (-0.95,-0.55) {};
    \node[childcheck] (rl) at (0.95,-0.55) {};
    \node[childcheck] (rr) at (2.15,-0.55) {};
    \node at (-1.55,-0.55) {$\cdots$};
    \node at (1.55,-0.55) {$\cdots$};
    \draw[msg] (ll) -- (left)
        node[midway,left=2pt] {$\boldsymbol q^{(\ell)}$};
    \draw[msg] (lr) -- (left);
    \draw[msg] (rl) -- (right);
    \draw[msg] (rr) -- (right)
        node[midway,right=2pt] {$\boldsymbol q^{(\ell)}$};
    \foreach \n in {ll,lr,rl,rr}
        \draw[densely dotted] (\n.south) -- +(0,-0.30);

    \node[draw,circle,minimum size=0.20cm,inner sep=0pt]
        at (-2.30,-1.20) {};
    \node[anchor=west,font=\footnotesize] at (-2.08,-1.20)
        {symbol: equality update};
    \node[childcheck] at (-2.30,-1.60) {};
    \node[anchor=west,font=\footnotesize] at (-2.08,-1.60)
        {check: parity update};
\end{tikzpicture}

\caption{One round of BPQM updates toward the root. Arrow labels give the classical parts of the messages; other children are omitted.}
\label{fig:qde-round}
\end{figure}
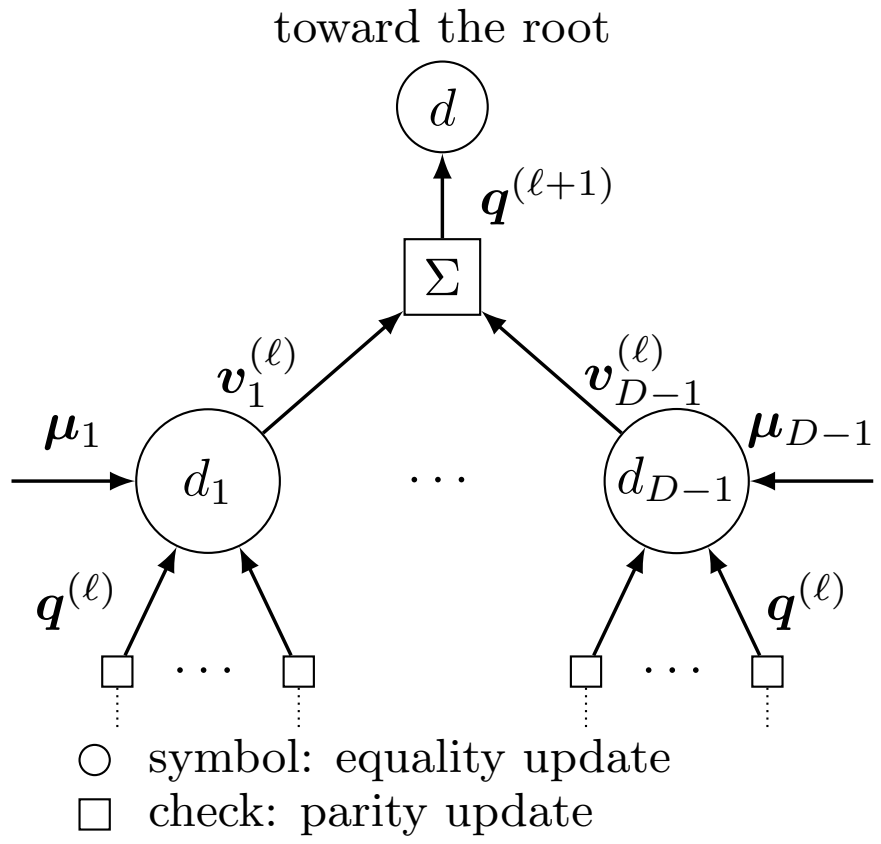

Let $\ell=0,1,2,\ldots$ index rounds toward the root. Write $\boldsymbol v^{(\ell)}$ and $\boldsymbol q^{(\ell)}$ for the classical parts of the symbol-to-check and check-to-symbol BPQM messages, respectively. One round is an equality update at the symbols followed by a parity update at their parent checks, as shown in \cref{fig:qde-round}.

The physical channel vector $\boldsymbol\mu_j$ may vary between symbols, as in \cref{eq:bpqm-initial}. We suppress node and edge indices in the local recursions below: $\boldsymbol\mu$ denotes the channel vector at the symbol being updated.

Numbering the incoming check messages, repeated application of the equality rule \cref{eq:bpqm-equality} gives the deterministic update
\begin{equation}
    \label{eq:qde-symbol}
    \boldsymbol v^{(\ell)}
        =\boldsymbol\mu*
          \boldsymbol q_1^{(\ell)}*\cdots*
          \boldsymbol q_{k-1}^{(\ell)}.
\end{equation}

For a check constraint $d=\sum_{t=1}^{D-1}\alpha_t d_t$, relabel each incoming message as in \cref{sec:bpqm}, setting $\nu_{t,a}^{(\ell)}=v_{t,\alpha_t a}^{(\ell)}$. Conditional on these incoming vectors, repeated application of the parity rule \cref{eq:bpqm-parity} gives a distribution over outgoing vector $\boldsymbol q^{(\ell+1)}$. In particular, we set $s_1=0$ and sample the string $\mathbf s=(s_2,\ldots,s_{D-1})\in\mathbb F_p^{D-2}$ with probability
\begin{equation}
    \label{eq:qde-outcomes}
    \Pr(\mathbf s)=Z_{\mathbf s}^{(\ell)}
        :=\sum_{a\in\mathbb F_p}
          \prod_{t=1}^{D-1}\nu_{t,a+s_t}^{(\ell)},
\end{equation}
then the associated outgoing vector
$\boldsymbol q^{(\ell+1)}$ is
\begin{equation}
    \label{eq:qde-check}
    q_a^{(\ell+1)}
        =\frac{\prod_{t=1}^{D-1}
                    \nu_{t,a+s_t}^{(\ell)}}{Z_{\mathbf s}^{(\ell)}},
    \qquad Z_{\mathbf s}^{(\ell)}>0.
\end{equation}

We initialize
\begin{equation}
    \label{eq:qde-initial}
    \boldsymbol q^{(0)}=(1,0,\ldots,0).
\end{equation}
Since $\boldsymbol\mu*\boldsymbol q^{(0)}=\boldsymbol\mu$, each initial symbol message has classical part $\boldsymbol v^{(0)}=\boldsymbol\mu$, using that symbol's physical channel vector. Let \(\mathcal Q_\ell\) denote the distribution of the classical part of a check-to-symbol message $\boldsymbol q^{(\ell)}$ under this ensemble. So the initialization gives \(\mathcal Q_0=\delta_{\boldsymbol e_0}\).

At the root, the equality rule combines the physical channel with all $k$ check messages. The classical part of the resulting message and the average decoding error are
\begin{equation}
    \label{eq:qde-root}
    \boldsymbol\mu_{\mathrm{root}}^{(\ell)}
        =\boldsymbol\mu*
          \boldsymbol q_1^{(\ell)}*\cdots*
          \boldsymbol q_k^{(\ell)},
    \qquad
    P_{\mathrm{err}}^{(\ell)}
        =\mathbb E\!\left[P_{\mathrm{err}}
            (\boldsymbol\mu_{\mathrm{root}}^{(\ell)})\right].
\end{equation}
Here $P_{\mathrm{err}}$ is given by \cref{eq:bpqm-error}, and the expectation averages over parity outcomes and, when applicable, the ensemble. The decoding threshold concerns whether $P_{\mathrm{err}}^{(\ell)}\to0$ as $\ell\to\infty$. In our numerical experiments, to determine the decoding threshold, we approximate the message distributions by collections of sampled vectors, apply these same updates, and average the root error.

\section{From Tree Decoding to Coherent Gibbs Sampling}
\label{sec:finite-sampling}

In \cref{sec:bpqm}, we described how BPQM estimates the root symbol on a tree, and \cref{sec:qde} recorded the recursion determining its error. This section describes how to decode the entire codeword on a finite Tanner graph, then uses that decoder in Regev's reduction to prepare the coherent Gibbs state.

\paragraph{Decoding the entire codeword.}
The lifting from tree BPQM decoding to global codeword decoding of random regular LDPC codes was established by Mandal, Piveteau, Renes, and Pfister \cite{MPRP26}. They consider a random bipartite graph with symbol degree $k$ and check degree $D$. Their ensemble allows parallel edges, unlike our ensemble in \cref{sec:ensemble}. Below, we for completeness briefly overview how the global codeword decoding algorithm works, and then adapt the result of~\cite{MPRP26} to our ensemble.

The lifting to global decoding~\cite{MPRP26} relies on the following well-known fact: up to any $\ell=\Theta(\log\log m)$, the radius-$\ell$ neighborhood of all but $o(m)$ symbols is a tree, with high probability over the graph ensemble. Their decoder applies BPQM sequentially on these trees, records each root-symbol estimate, and reverses the BPQM unitaries. Recall that for use in Regev's reduction, the measurement outcomes are retained coherently. A quantum union bound controls the accumulated measurement disturbance, and because the tree BPQM decoding error, if converging to zero, can be shown to decay doubly exponentially with the tree depth $\ell$~\cite{MPRP26},  the total error of the BPQM stage vanishes. Symbols with non-tree neighborhoods are then treated as erasures and recovered by Gaussian elimination. There are only $o(m)$ such symbols, while the code's minimum distance is linear in $m$, so erasure recovery succeeds provided the BPQM estimates are correct. Erasure correction is done by Gaussian elimination, implemented reversibly.

We now specialize this result to our code ensemble and to the physical channel of \cref{eq:bpqm-message}.

\begin{proposition}[Global decoding from tree decoding~\cite{MPRP26}]
    \label{thm:finite-bpqm}
    Let $(B,\vect F)$ be drawn from the ensemble $\GG_{\reg}(n,k,D,p,S,\Lambda)$ of \cref{sec:ensemble}. At any fixed $\beta\ge 0$, let $\boldsymbol\mu$ be the channel vector in \cref{eq:bpqm-initial} for the common target set $S$. Suppose the recurrence in \cref{sec:qde} satisfies
    $\mathbb E[P_{\mathrm{err}}(\boldsymbol v^{(\ell)})]\to 0.$
    Then, for a suitable $\ell=\Theta(\log\log m)$, the BPQM decoder with erasure recovery of \cite[Theorem~30]{MPRP26} runs in polynomial time and decodes a uniformly chosen codeword of $C^\perp=\ker B^T$ with ensemble-average error probability tending to zero as $m$ increases.
\end{proposition}

\begin{proof}
The result follows from \cite[Theorem~30]{MPRP26} with two differences to address: we allow a more general coefficient distribution $\Lambda$, and we disallow parallel edges. Their code has $m$ symbols, symbol degree $k$, check degree $D$, and alphabet size $p$ in our notation.

\emph{Coefficient distribution.}
First use the random graph of \cite{MPRP26}, generated by uniformly matching the $mk=nD$ edge ends on each side of the bipartite Tanner graph, and independently assign a coefficient tuple with law $\Lambda$ to the $D$ ends at each check. Although the coefficients in \cite{MPRP26} are independent and uniform, the combining bounds in their Lemmas~24--25 hold for every choice of nonzero coefficients, with right-hand sides independent of those coefficients. Condition on a check's entire tuple: its incoming messages depend on disjoint descendant subtrees, which use independent column tuples and are independent of the conditioned tuple. Thus correlations within a column do not affect the averaging step in their proof of Theorem~26. Together with their Lemma~18, that argument still gives doubly-exponential-in-$\ell$ decay of the tree decoding error, under the assumption $\mathbb E[P_{\mathrm{err}}(\boldsymbol v^{(\ell)})]\longrightarrow0.$

The linear code distance input to their Theorem~30 also holds for our coefficient law. A nonzero codeword's support has no check incident to exactly one of its edges. A set of $s$ symbols with this property reaches at most $\lfloor ks/2\rfloor$ checks. We take a union bound over the choice of the $s$ symbols and of $\lfloor ks/2\rfloor$ checks containing all their neighbors; the probability that all $ks$ edge ends of these symbols are matched into those checks is at most $(\lfloor ks/2\rfloor/n)^{ks}$. For sufficiently small fixed $\delta>0$, this gives
\begin{align*}
    \Pr\{\text{some nonempty such set has size at most $\delta m$}\}
    &\leq\sum_{s=1}^{\lfloor\delta m\rfloor}
    \binom{m}{s}\binom{n}{\lfloor ks/2\rfloor}\left(\frac{\lfloor ks/2\rfloor D}{n D}\right)^{ks}\\
    &\leq\sum_{s=1}^{\lfloor\delta m\rfloor}
    \left[e\left(\frac{eD}{2}\right)^{k/2}\left(\frac{s}{m}\right)^{k/2-1}\right]^s=o(1),
\end{align*}
where the second inequality uses $\binom{N}{i}\leq(eN/i)^i$ and $n=mk/D$.
Hence the minimum distance exceeds $\delta m$ with high probability, regardless of the nonzero coefficients. The local decoding, cycle counting, and erasure-recovery arguments in their Lemmas~21, 27--28 and Theorem~30 now apply unchanged. In particular, let $\varepsilon(G)$ denote the decoder's block-error probability for a graph $G$ drawn by the matching construction of \cite[Sec.~4.1]{MPRP26}, with coefficient law $\Lambda$ as above. These arguments give $\mathbb E_G[\varepsilon(G)]\to0$ at a suitable $\ell=\Theta(\log\log m)$.

\emph{No parallel edges.}
Let $\mathcal A$ be the event that the matched graph has no parallel edges. Each labeled simple $(k,D)$-regular bipartite graph arises from exactly $(k!)^m(D!)^n$ matchings, since the edge ends at each vertex can be assigned to its incident edges in any order. Conditional on $\mathcal A$, the support is therefore uniform among the graphs in \cref{sec:ensemble}. The column tuples remain independent because $\mathcal A$ depends only on the matching, and exchangeability preserves their law $\Lambda$ when ordered by neighboring symbols. This gives precisely our ensemble, including both $\Lambda_{\rm iid}$ and $\Lambda_{\rm perm}$.

For fixed $k,D$, $\liminf_{m\to\infty}\Pr(\mathcal A)>0$ \cite[Theorem~5.1]{J13}. Writing $\varepsilon_B$ for the decoder's block-error probability on our ensemble, we conclude
\begin{equation}
    \label{eq:finite-simple}
    \mathbb E_B[\varepsilon_B]
       =\mathbb E_G[\varepsilon(G)\mid\mathcal A]
       \leq\frac{\mathbb E_G[\varepsilon(G)]}{\Pr(\mathcal A)}
       \longrightarrow0.
\end{equation}
\end{proof}

\paragraph{Coherent Gibbs sampling.}
We now combine the global quantum decoder and Regev's reduction to perform coherent Gibbs sampling for the ensemble of \cref{sec:ensemble}. Recall that each target set $F_j=\tau_j+S$ is a random translate of a fixed set $S$. We use the same decoder for every choice of shifts: it decodes the channel associated with $S$, while the shifts $\tau_j$ are implemented by phase gates in Regev's reduction. According to~\eqref{eq:bpqm-initial}, the unshifted channel has parameters
\begin{equation}
    \label{eq:finite-channel}
    \mu_a=\frac{1}{r+(p-r)e^{-\beta}}
    \begin{cases}
        1,&a\in S,\\
        e^{-\beta},&a\notin S,
    \end{cases}
    \qquad r=|S|.
\end{equation}
We will apply \cref{thm:finite-bpqm} to solve the quantum decoding problem associated to this channel.
The random shifts $\tau_j$ allow us to apply \cite[Lemma~4]{SMR26} to convert quantum decoding error to Regev's reduction error.

\begin{corollary}[Coherent Gibbs sampling]
    \label{cor:finite-sampling}
    Let $(B,\vect F)\sim \GG_{\reg}(n,k,D,p,S,\Lambda)$ as in \cref{sec:ensemble}. Fix an inverse temperature $\beta\geq0$. Suppose the check-to-symbol messages $\boldsymbol q^{(\ell)}$ of the density evolution in \cref{sec:qde}, for this law and the channel \cref{eq:finite-channel}, satisfy $\mathbb E\big[\|\boldsymbol q^{(\ell)}-\boldsymbol u\|_2^2\big]\to0$, where $\boldsymbol u=(1/p,\ldots,1/p)$.
    Then, with probability $1-o(1)$ over $(B,\boldsymbol\tau)$, Regev's reduction with the decoder of \cref{thm:finite-bpqm} prepares a state within $o(1)$ trace distance of the coherent Gibbs state $\ket{\psi_\beta}$ in \cref{eq:psi_Gibbs}, in polynomial time and with failure probability $o(1)$, as $n\to\infty$. This applies to both $\Lambda_{\rm iid}$ and $\Lambda_{\rm perm}$.
\end{corollary}

\begin{proof}
The hypothesis and \cref{eq:qde-symbol} imply $\mathbb E[P_{\mathrm{err}}(\boldsymbol v^{(\ell)})]\to0$. By \cref{thm:finite-bpqm}, the decoder's block-error probability $\varepsilon_B$ satisfies $\mathbb E_B[\varepsilon_B] \to0$.

Let $\ket{v_{\mathbf d}}$ be the normalized received state \cref{eq:corrupted-codeword} with every target set equal to $S$. Write the coherent decoder's action as
\[
    U_{\mathrm{dec}}\bigl(\ket{v_{\mathbf d}}\ket{\mathbf0}\bigr)
       =\sum_{\widehat{\mathbf d}\in\mathbb F_p^m}
          \ket{\eta_{\widehat{\mathbf d},\mathbf d}}\ket{\widehat{\mathbf d}},
\]
where each unnormalized $\ket{\eta_{\widehat{\mathbf d},\mathbf d}}$ contains the received system and all other decoder registers. Following \cite[Sec.~7.2]{SMR26}, apply the phase $ \omega^{-\langle\boldsymbol\tau,\mathbf d\rangle}$, apply the decoding unitary $U_{\mathrm{dec}}$, and test whether the estimate $\widehat{\mathbf d}$ agrees with register $\mathbf d$. Report failure otherwise. The test accepts with probability $1-\varepsilon_B$. On the accepting branch, subtracting the estimate from the codeword register gives
\[
    \frac1{\sqrt {|C^\perp|}}\sum_{\mathbf d\in C^\perp}
       \omega^{-\langle\boldsymbol\tau,\mathbf d\rangle}
       \ket{\mathbf d}\ket{v_{\mathbf d}}\ket{\mathbf0}\xrightarrow{\text{decode, test, subtract}}
       \ket{\mathbf0}\otimes
       \frac{\displaystyle\sum_{\mathbf d\in C^\perp}
          \omega^{-\langle\boldsymbol\tau,\mathbf d\rangle}
          \ket{\eta_{\mathbf d,\mathbf d}}\ket{\mathbf d}}
          {\sqrt{|C^\perp|(1-\varepsilon_B)}}.
\]
Now reverse $U_{\mathrm{dec}}$, apply $Z_{\boldsymbol\tau}\ket{\mathbf y}=\omega^{\langle\boldsymbol\tau,\mathbf y\rangle}\ket{\mathbf y}$ to the received register, and apply the inverse Fourier transform. To see what these steps do, consider an ideal decoder, $\varepsilon_B=0$, for which reversing $U_{\mathrm{dec}}$ maps $\ket{\eta_{\mathbf d,\mathbf d}}\ket{\mathbf d}$ back to $\ket{v_{\mathbf d}}\ket{\mathbf 0}$. Write $P^S_\beta$ for the envelope \cref{eq:P1} with every $F_j$ replaced by $S$, so that $\ket{v_{\mathbf d}}=\|P^S_\beta\|_2^{-1}\sum_{\mathbf e}\widetilde P^S_\beta(\mathbf e)\ket{\mathbf d+\mathbf e}$. Since $F_j=\tau_j+S$, \cref{eq:QFT} gives
\[
    \frac{\widetilde P_\beta(\mathbf e)}{\|P_\beta\|_2}
    =\omega^{\langle\boldsymbol\tau,\mathbf e\rangle}
      \frac{\widetilde P^S_\beta(\mathbf e)}{\|P^S_\beta\|_2}.
\]
Omitting the work registers,
\[
\begin{aligned}
    \frac{1}{\sqrt{|C^\perp|}}\sum_{\mathbf d\in C^\perp}
       \omega^{-\langle\boldsymbol\tau,\mathbf d\rangle}Z_{\boldsymbol\tau}\ket{v_{\mathbf d}}
    &=\frac{1}{\sqrt{|C^\perp|}\,\|P^S_\beta\|_2}\sum_{\mathbf d\in C^\perp}\sum_{\mathbf e\in\mathbb F_p^m}
       \omega^{\langle\boldsymbol\tau,\mathbf d+\mathbf e\rangle-\langle\boldsymbol\tau,\mathbf d\rangle}\,
       \widetilde P^S_\beta(\mathbf e)\ket{\mathbf d+\mathbf e}\\
    &=\frac{1}{\sqrt{|C^\perp|}\,\|P_\beta\|_2}\sum_{\mathbf d\in C^\perp}\sum_{\mathbf e\in\mathbb F_p^m}
       \widetilde P_\beta(\mathbf e)\ket{\mathbf d+\mathbf e},
\end{aligned}
\]
Up to normalization, this is $\ket{\widetilde C_\beta}$ in \cref{eq:noisy-dual-codeword}, so the inverse Fourier transform gives the target state $\ket{\psi_\beta}$. For nonzero $\varepsilon_B$, denote the normalized output, including all work registers, by $\ket{\Psi_{B,\boldsymbol\tau}}$.  The error analysis of \cite[Lemma~4]{SMR26} gives
\begin{equation}
    \label{eq:finite-coherent-fidelity}
    \mathbb E_{\boldsymbol\tau}\!\left[
       1-\left|\bigl(\bra{\psi_\beta}\otimes\bra{\mathbf0}_{\mathrm{work}}\bigr)
            \ket{\Psi_{B,\boldsymbol\tau}}\right|^2
    \right]\leq\varepsilon_B.
\end{equation}
The left-hand side is the expected squared trace distance between these pure states. Averaging over $B$ and applying Markov's inequality gives trace-distance error and failure probability $o(1)$, with probability $1-o(1)$ over $(B,\boldsymbol\tau)$.
\end{proof}

In \cref{sec:main-theorem}, we will show that the convergence condition in \cref{cor:finite-sampling} corresponds precisely to the reconstruction threshold on the inverse temperature $\beta$.

\section{The Reconstruction Threshold}
\label{sec:reconstruction}

In many random optimization problems, spin glass theory predicts a dynamical transition that occurs at a critical inverse temperature $\beta_{\mathrm{dyn}}$, which is determined by when there exists a nontrivial solution to the one-step replica-symmetry-breaking (1RSB) equation with Parisi parameter $m=1$~\cite{Montanari2003,MM06}. According to M\'ezard and Montanari~\cite{MM06}, these equations can be mapped to a reconstruction problem on the infinite tree. We use this reconstruction problem to define $\beta_{\mathrm{dyn}}$ below.

\paragraph{The Reconstruction Problem.}
Following \cite{MM06}, we introduce the reconstruction problem as a communication problem on a tree. A source at the root chooses a symbol uniformly from $\mathbb F_p$ and sends it through noisy channels to its children. Each child then transmits to the next generation in the same way. After $\ell$ generations, an observer sees only the symbols at that last generation. The task is to infer the original root symbol from these observations. Reconstruction is possible if, as $\ell\to\infty$, the observer can still guess the root with a success probability that exceeds $1/p$ by a fixed positive amount.

For our spin model, the tree has variable nodes and constraint nodes, just as in the Tanner graph of \cref{sec:ensemble}. Each variable belongs to $D$ constraints, and each constraint contains $k$ variables. A constraint acts as a channel from one parent variable to $k-1$ child variables, whose values are generated jointly. This is the one-to-many broadcast considered in \cite[Section 7]{MM06}. One generation passes through a constraint to its child variables. The observer knows the tree, its coefficients, and its target offsets; we denote this information by $\mathcal T$. 

We now specify the channel for which the broadcast generates our desired Gibbs distribution. Fix a nonempty proper accepting set $S\subsetneq\mathbb F_p$, let $r=|S|$, and write
\begin{equation}
    \label{eq:reconstruction-weight}
    w_\beta(u)=e^{-\beta}+(1-e^{-\beta})\mathbf1\{u\in S\},
    \qquad c_\beta=r+(p-r)e^{-\beta},
    \qquad \mu_a=\frac{w_\beta(a)}{c_\beta}.
\end{equation}
For a constraint $j$ with parent variable $i$ and children $i_1,\ldots,i_{k-1}$, the probability of the child values, given the parent, is
\begin{equation}
    \label{eq:reconstruction-kernel}
    K_j(x_{i_1},\ldots,x_{i_{k-1}}\mid x_i)
      =\frac{w_\beta\!\left(B_{ji}x_i+\sum_{h=1}^{k-1}B_{ji_h}x_{i_h}-\tau_j\right)}
             {p^{k-2}c_\beta}.
\end{equation}
To sample it, draw a residual $U_j\sim\boldsymbol\mu$, then choose the $k-1$ children uniformly subject to
$B_{ji}x_i+\sum_h B_{ji_h}x_{i_h}-\tau_j=U_j$.
All coefficients are nonzero, so there are exactly $p^{k-2}$ choices. Start with a uniform root and apply this rule at each constraint, using independent randomness at different constraints. The probability of a complete assignment is then proportional to $e^{-\beta E(\mathbf x)}$: it is precisely the Gibbs distribution on the finite tree, with no spins fixed at the boundary. This is called the \emph{free-boundary Gibbs measure}. Continuing the broadcast defines it on the infinite tree as well. Note that children of one constraint can be correlated, but their subsequent broadcasts are independent once their values and $\mathcal T$ are fixed.

\begin{definition}[Reconstruction threshold]
    For the broadcast \cref{eq:reconstruction-kernel}, let $X_0$ be the uniform root symbol and $X_{\partial T_\ell}$ the symbols observed at generation $\ell$. The conditional mutual information retained about the root is
    \begin{equation}
        \label{eq:reconstruction-definition}
        I_\ell(\beta):=I(X_0;X_{\partial T_\ell}\mid\mathcal T).
    \end{equation}
    Reconstruction is possible at $\beta$ if $\lim_{\ell\to\infty}I_\ell(\beta)>0$. For a fixed accepting set and coefficient law, the reconstruction threshold is
    \begin{equation}\label{eq:linsat-betadyn}
       \displaystyle
        \beta_{\dyn}:=\inf\{\beta>0:\lim_{\ell\to\infty}I_\ell(\beta)>0\},
    \end{equation}
    with $\inf\varnothing=\infty$.
\end{definition}

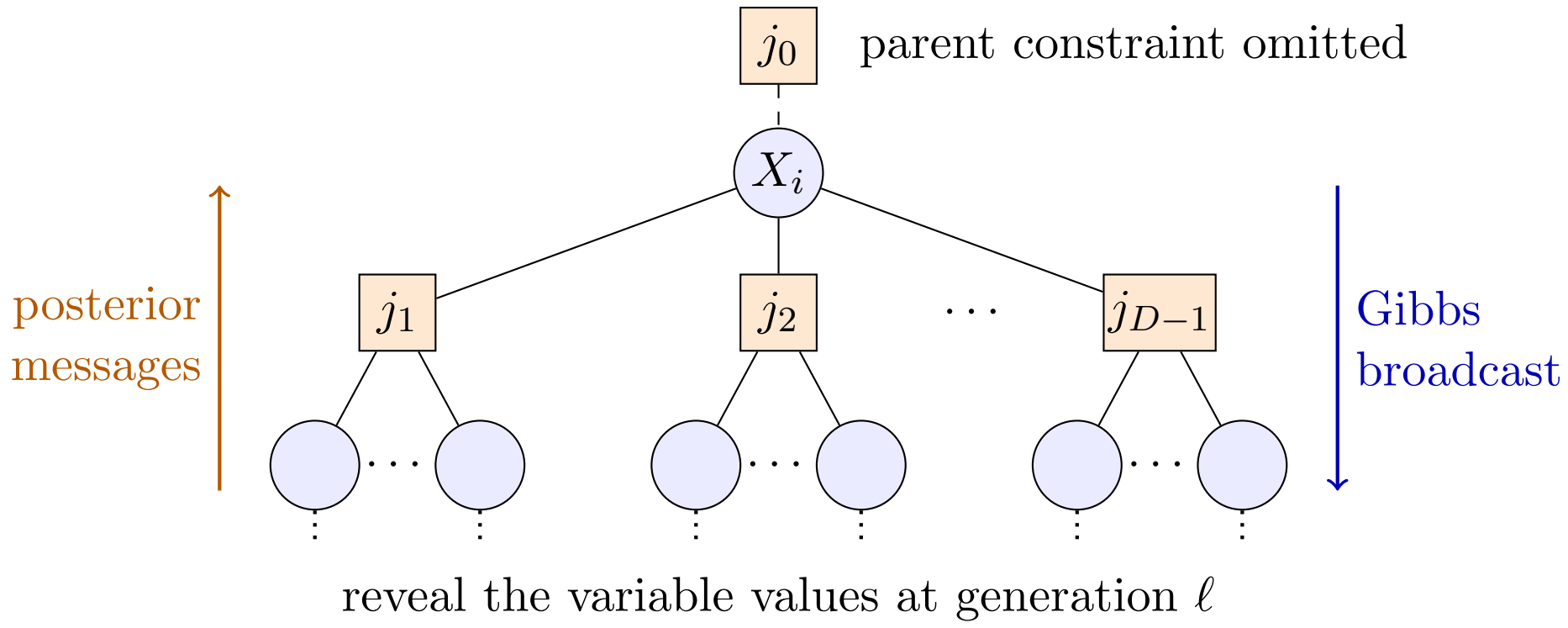
\begin{figure}[t]
\centering
\begin{tikzpicture}[x=1cm,y=1cm,
  spin/.style={circle,draw,fill=blue!8,minimum size=7mm,inner sep=1pt},
  fac/.style={rectangle,draw,fill=orange!18,minimum size=6mm,inner sep=1pt}]
  \node[fac] (parent) at (0,0.8) {$j_0$};
  \node[spin] (root) at (0,-0.2) {$X_i$};
  \draw[dashed] (parent)--(root);
  \node[anchor=west] at (0.5,0.8) {parent constraint omitted};
  \foreach \t/\xx/\lab in {1/-3/1,2/0/2,3/3/{D-1}} {
    \node[fac] (f\t) at (\xx,-1.3) {$j_{\lab}$};
    \draw (root)--(f\t);
    \node[spin] (a\t) at (\xx-0.65,-2.5) {};
    \node[spin] (b\t) at (\xx+0.65,-2.5) {};
    \draw (f\t)--(a\t);
    \draw (f\t)--(b\t);
    \node at (\xx,-2.5) {$\cdots$};
    \draw[dotted,thick] (a\t)--(\xx-0.65,-3.1);
    \draw[dotted,thick] (b\t)--(\xx+0.65,-3.1);
  }
  \node at (1.55,-1.3) {$\cdots$};
  \node at (0,-3.55) {reveal the variable values at generation $\ell$};
  \draw[->,thick,blue!70!black] (4.4,-0.3)--(4.4,-2.7)
      node[midway,right,align=left] {Gibbs\\broadcast};
  \draw[->,thick,orange!70!black] (-4.4,-2.7)--(-4.4,-0.3)
      node[midway,left,align=right] {posterior\\messages};
\end{tikzpicture}
\caption{The reconstruction problem. Each of the $D-1$ child constraints jointly generates $k-1$ child spins using \cref{eq:reconstruction-kernel}. After broadcasting to depth $\ell$, reveal the boundary and compute the posterior of $X_i$ by passing messages inward. The full-root problem includes all $D$ constraints. Under transposition to the decoding graph, circles become parity checks and squares become code symbols with physical channel outputs.}
\label{fig:reconstruction-tree}
\end{figure}

At $\beta=0$, the children are uniform and independent of the parent, so the boundary carries no information about the root. Increasing $\beta$ makes the constraints more informative. Thus reconstruction fails below $\beta_{\mathrm{dyn}}$ and succeeds above it.

\paragraph{Reconstruction threshold from posterior distributional recursion.}
We now describe the recursion equation for the posterior distribution for any specified $(k,D,p,S,\Lambda)$, and in turn provide a more explicit definition of $\beta_\dyn$.
The recursion is performed on a probability vector $\widetilde{\boldsymbol\eta}$ on $\mathbb F_p$, which we call a reconstruction message, whose entry $\widetilde\eta(z)$ denotes the posterior probability that the root differs from its ``planted value'' by $z$, after rescaling by the coefficient on its parent constraint. 
More precisely, we draw a reference configuration $\xv^*$ from the Gibbs broadcast so that $x_i^*$ is the planted value at the root node $i$; if the inferred root variable is $x_i$ and its parent constraint is $j_0$, then the error coordinate is $z=B_{j_0i}(x_i-x_i^*)$. 
For its child constraints $j_t$, write $\gamma_t=B_{j_ti}/B_{j_0i}\in\mathbb F_p^\times$, with $t=1,\ldots,D-1$.

Let $\widetilde{\boldsymbol\eta}^{(0)}=\boldsymbol e_0$, the point mass at $z=0$, corresponding to zero error at the $\ell=0$ generation.
For each $\ell\geq0$, draw one $D$-tuple of coefficients from the given law $\Lambda$ and compute the coefficient ratios $\gamma_1,\ldots,\gamma_{D-1}$ of the $D-1$ children relative to their parent.
Here, for each $t\in [D-1]$, we choose the noise residual $U_t \sim^\iid \boldsymbol{\mu}$ from \eqref{eq:reconstruction-weight}, and independently draw the $k-1$ copies of $\widetilde{\boldsymbol\eta}^{(\ell)}$ as the child messages to obtain
\begin{equation}\label{eq:reconstruction-convolution}
    \boldsymbol G_t :=
        \widetilde{\boldsymbol\eta}^{(\ell)}_{t,1}*\cdots*
        \widetilde{\boldsymbol\eta}^{(\ell)}_{t,k-1},
    \qquad
    \text{and} \qquad
    \Gamma_t(s):= \Pr_{Z\sim G_t} [Z+s\in S] =\sum_{v\in S}G_t(v-s).
\end{equation}
where $*$ is the circular convolution in \cref{eq:bpqm-equality}. 
The interpretation here is that the convolution $G$ is the distribution of the sum of the child errors in the constraint's coordinates; summing the Gibbs weight over these errors gives the likelihood $\sum_w G_t(w)w_\beta(U_t+\gamma_tz+w) = e^{-\beta} + (1-e^{-\beta})\Gamma_t(U_t+\gamma_t z)$.
Then the posterior message distribution at generation $\ell+1$ is
\begin{equation}\label{eq:app-linsat-de}
    \widetilde\eta^{(\ell+1)}(z)\overset{d}{=}
    \frac{1}{\mathcal N}\prod_{t=1}^{D-1}
    \Big[e^{-\beta}+(1-e^{-\beta})\,
        \Gamma_t(U_t+\gamma_t z)\Big],
    \qquad z\in\mathbb F_p.
\end{equation}
where $\mathcal{N}$ is a normalization fixed by $\sum_z\tilde\eta^{(\ell+1)}(z)=1$.

To measure the information in the posterior about the root variable, we compare it against the uniform random guess by defining
\begin{equation}\label{eq:reconstruction-order}
    \Psi_\ell:=\mathbb E[\widetilde\eta^{(\ell)}(0)]-\frac1p,
    \qquad
    \Psi_\infty(\beta;k,D,p,S,\Lambda):=\lim_{\ell\to\infty}\Psi_\ell.
\end{equation}
Note $\Psi_\ell$ is the excess success probability of guessing the root correctly by sampling from the posterior; in particular, $\Psi_0=1-1/p$.

Since a farther boundary is obtained by continuing the broadcast from a nearer one,  $\Psi_\ell$ is nonincreasing in $\ell$ due to the data processing inequality.
As noted in \cite{MM06}, the limit $\Psi_\infty$ exists and vanishes exactly when the mutual information vanishes.
Thus, the reconstruction threshold \eqref{eq:linsat-betadyn} can be equivalently written as
\begin{equation}\label{eq:linsat-betadyn-explicit}
    \beta_{\mathrm{dyn}}(k,D,p,S,\Lambda)
       :=\inf\{\beta>0:\ \Psi_\infty(\beta;k,D,p,S,\Lambda)>0\} 
\vphantom{\Big|}
.
\end{equation}

\paragraph{Remarks.}
\begin{enumerate}[label=(\roman*)]
\item We remark that the reconstruction information $\Psi_\ell(\beta)$ is nondecreasing in $\beta$, making $\beta_\dyn$ a genuine threshold. This enables us to locate the transition $\beta_\dyn$ by scanning $\beta$ and bisecting. In \cref{cor:beta-monotonicity} of \cref{sec:main-theorem}, we give a proof of this monotonicity using channel duality.
\item The formulation above recovers known recursions for Ising model/max-XORSAT as a special case (see e.g., \cite{ZSSJ26}). At $(p,r)=(2,1)$, write $M_\ell=\widetilde\eta^{(\ell)}(0)-\widetilde\eta^{(\ell)}(1)$ for the posterior magnetization. Then \cref{eq:app-linsat-de} reduces to the max-XORSAT recursion in \cite[Appendix~D.2]{ZSSJ26}:
\begin{equation}\label{eq:xor-de}
    M_{\ell+1}\overset{d}{=}
    \tanh\!\left[\sum_{t=1}^{D-1}
        \operatorname{artanh}\!\left(
            \eta\,\sigma_t\prod_{j=1}^{k-1}M_{\ell,t,j}
        \right)\right].
\end{equation}
Here $\eta=\tanh(\beta/2)$, the children are independent copies of $M_\ell$, and $\sigma_t=(-1)^{U_t}$ are independent signs with $\Pr(\sigma_t=s)=(1+s\eta)/2$. In particular, $\Psi_\ell=\mathbb E[M_\ell]/2$.

\item When the accepting set size is $|S|=1$ or $p-1$, the threshold $\beta_\dyn(k,D,p,S,\Lambda)$ is in fact independent of $\Lambda$. 
This is because on the tree, a change of variables exists, via $x_i=a_i y_i + b_i$, so that each constraint $\sum_i B_{ji} x_i -\tau_j \in S$ becomes either $\sum_{i\in \partial j} y_i = 0$ or $\sum_{i\in \partial j} y_i \neq 0$, when $|S|=1$ or $p-1$ respectively.
For other accepting-set sizes, however, $\beta_\dyn$ may depend on the choice of $\Lambda$.

\item If instead of $\GG_\reg$, one is interested in the Erd\H{o}s--R\'enyi ensemble of $k$-uniform hypergraphs with average degree $D$, the corresponding tree recursion replaces $D-1$ in \cref{eq:app-linsat-de} by an independently random offspring count $d\sim\operatorname{Poisson}(D)$. Then one needs to specify a family of exchangeable, degree-dependent coefficient law $\Lambda$, and draw $d+1$ coefficients for the clauses incident to every variable node.
\end{enumerate}

\paragraph{Connection to BPQM.}
The reconstruction distributional recursion resembles quantum density evolution in \cref{sec:qde}, as both pass probability vectors inward on the same tree. Their information meanings are reversed, however: a uniform classical posterior means that the root cannot be inferred, whereas a uniform BPQM vector describes orthogonal quantum states and permits perfect decoding. The next section explains this correspondence and proves the two thresholds coincide.


\section{Proof of the Main Theorem}
\label{sec:main-theorem}

This section proves our main theorem: the BPQM decoding threshold coincides with the reconstruction threshold. 
The resemblance between the BPQM and reconstruction recursions on the infinite tree becomes an exact correspondence after a random translation and reflection of each BPQM vector. The correspondence can be understood via Renes' channel duality~\cite{R18}, which explains why impossibility of reconstruction coincides with decodability with vanishing error.
For our channels, this duality reduces to simple identities for sums and conditional probabilities which we verify directly, after which we relate the message convergence to the root decoding error.
Combining this coincidence with \cref{cor:finite-sampling}, we also show Regev+BPQM can coherently Gibbs sample the corresponding max-LINSAT Hamiltonian up to the reconstruction threshold, which is the conjectured limit on classical local dynamics from the dynamical glass transition.

Fix any nonempty proper accepting set $S \subsetneq \FF_p$ and a coefficient law $\Lambda$ for both recursions.
Let $\mathcal Q_\ell$ be the BPQM check-to-symbol message law in \cref{sec:qde}, using the unshifted physical channel $\boldsymbol\mu$ of \cref{eq:finite-channel}, and let $P_{\mathrm{err}}^{(\ell)}$ be its ensemble-average terminal error from \cref{eq:qde-root}. Define
\begin{equation}
    \label{eq:bpqm-threshold}
    \beta_{\mathrm{BPQM}}(k,D,p,S,\Lambda)
       :=\inf\left\{\beta>0:\
           \lim_{\ell\to\infty}P_{\mathrm{err}}^{(\ell)}>0\right\}.
\end{equation}
Note the error limit exists because increasing the tree depth supplies more channel outputs and BPQM achieves the minimum root error on each tree~\cite{PR22}. We restate \cref{thm:main-informal} formally:

\begin{samepage}
\begin{theorem}[Regev+BPQM reaches reconstruction threshold]
\label{thm:main}
    Fix a prime $p$, integers $3\leq k<D$, and a nonempty proper subset $S\subsetneq\mathbb F_p$. 
    At each optimization variable, independently draw its incident nonzero coefficients from an exchangeable law $\Lambda$ on $(\mathbb F_p^\times)^D$. Use this same law for reconstruction and BPQM recursions.
    For a probability vector $\boldsymbol q$ over $\mathbb F_p$, define its random posterior lift vector by
    \begin{equation}
    \label{eq:qde-posterior-lift}
        \mathcal L(\boldsymbol q)_z :=q_{A-z},
        \qquad \text{with} \qquad A \sim\boldsymbol q.
    \end{equation}
    Let $\boldsymbol q^{(0)}=\widetilde{\boldsymbol\eta}^{(0)}=\boldsymbol e_0$, $\boldsymbol q^{(\ell)}\sim\mathcal Q_\ell$, and $\widetilde{\boldsymbol\eta}^{(\ell)}$ be as in \eqref{eq:app-linsat-de}.
    Then, for every finite $\beta\geq0$ and depth $\ell\geq0$,
    \begin{equation}
        \label{eq:main-finite-depth}
        \widetilde{\boldsymbol\eta}^{(\ell)}
             \overset{d}{=}\mathcal L(\boldsymbol q^{(\ell)}),
        \qquad
        \Psi_\ell=\mathbb E\!\left[
            \|\boldsymbol q^{(\ell)}-\boldsymbol u\|_2^2\right],
    \end{equation}
    where $\boldsymbol u=(1/p,\ldots,1/p)$. Their thresholds satisfy
    \begin{equation}
        \label{eq:main-threshold-equality}
        \beta_{\mathrm{BPQM}}(k,D,p,S,\Lambda)
           =\beta_{\mathrm{dyn}}(k,D,p,S,\Lambda).
    \end{equation}
For the finite-graph ensemble of \cref{sec:ensemble}, every fixed $0\leq\beta<\beta_{\mathrm{dyn}}$ admits polynomial-time coherent Gibbs sampling by Regev+BPQM, with the error and failure guarantees of \cref{cor:finite-sampling}.
\end{theorem}
\end{samepage}

We remark that the threshold equality \eqref{eq:main-threshold-equality} concerns the ideal infinite tree recursions. The result does not establish any shattering or provable hardness of sampling on finite graphs at $\beta> \beta_\dyn$, which requires additional arguments (see e.g., \cite{Alaoui2025sampling,alaoui2026near,ZSSJ26}).

\paragraph{Channel duality.}
To prepare for the proof, we first establish two identities \eqref{eq:qde-dual-convolution} and \eqref{eq:qde-dual-product} arising from Renes' channel duality~\cite{R18}, which we will use to connect the update rules for two recursions.
This channel duality will also explain the apparent reversal in the meaning of information in the two settings: \emph{Near-perfect BPQM decoding is possible if and only if reconstruction is impossible.}

Write $W_{\boldsymbol q}$ for the BPQM message channel $d\mapsto\ket{\phi_d(\boldsymbol q)}$ in \cref{eq:bpqm-message}. Its dual is equivalent, up to a fixed basis change and relabeling, to the classical additive channel
\begin{equation}
    \label{eq:qde-dual-channel}
    W_{\boldsymbol q}^{\perp}(y\mid x)=q_{y-x},
    \qquad \text{i.e., }\quad  Y=X+A,\quad A\sim\boldsymbol q,
\end{equation}
where $X$ is the input whose prior distribution we will take to be uniform on $\mathbb F_p$. Renes' duality theorem exchanges equality and parity combinations for binary-alphabet channels \cite[Theorem~1]{R18}. For our channel family over $\mathbb F_p$, the same exchange follows from the posterior representation that we verify below.

With the uniform prior $X \sim \mathbb{F}_p$, the posterior after observing $Y=y$ has entries $\mathrm{Pr}(X = x | Y = y) = q_{y-x}$. Sampling observations conditional on the transmitted value $X=0$, such that $Y=A$, gives precisely the random posterior lift $\mathcal L(\boldsymbol q)_x=q_{A-x}$ defined in \eqref{eq:qde-posterior-lift}.
Two independent such posteriors combine by convolution when estimating the sum of their symbols, and by normalized component-wise multiplication when they are independent observations of the same symbol. Specifically,
consider any two probability vectors $\boldsymbol q$ and $\boldsymbol r$ on $\FF_p$, and let $\boldsymbol w$ be their parity output as in \cref{eq:bpqm-parity}: draw $s$ with probability $\pi_s=\sum_a q_a r_{a+s}$ and set $w_a=q_a r_{a+s}/\pi_s$. Accordingly,
\begin{align}
    \mathcal L(\boldsymbol q*\boldsymbol r)
        &\overset{d}{=}\mathcal L(\boldsymbol q)*\mathcal L(\boldsymbol r),
          \label{eq:qde-dual-convolution}\\
    \mathcal L(\boldsymbol w)_x
        &\overset{d}{=}
        \frac{\mathcal L(\boldsymbol q)_x\mathcal L(\boldsymbol r)_x}
             {\sum_{y\in\mathbb F_p}
                \mathcal L(\boldsymbol q)_y\mathcal L(\boldsymbol r)_y}.
    \label{eq:qde-dual-product}
\end{align}
These are identities of joint vector distributions, with independent input lifts on each right-hand side. For the first, use independent noises $A_1\sim\boldsymbol q$ and $A_2\sim\boldsymbol r$ and add them. For the second, set $s=A_2-A_1$. Its probability and the conditional distribution of $A_1$ are exactly the branch probability and outgoing vector in \cref{eq:bpqm-parity}. Evaluating the outgoing vector at $a=A_1-x$ gives the normalized product in \cref{eq:qde-dual-product}. For the coefficient relabeling in \cref{sec:qde}, write $(\mathcal R_\alpha\boldsymbol q)_a=q_{\alpha a}$; then $\mathcal L(\mathcal R_\alpha\boldsymbol q)_x\overset{d}{=}\mathcal L(\boldsymbol q)_{\alpha x}$. Thus quantum symbol updates become classical check updates, and quantum check updates become classical symbol updates under Renes' channel duality.

Furthermore, we remark that Renes' tight entropic uncertainty relation for Fourier-conjugate inputs gives complementary information quantities \cite[Theorem~2 and Corollary~4]{R18}, making the information reversal between the two recursions quantitative.
With $H(\boldsymbol q)=-\sum_a q_a\log q_a$, for a fixed probability vector $\boldsymbol q$ and uniform input, the quantum Holevo information $\chi$ of $W_{\boldsymbol q}$ and the classical mutual information $I$ of $W_{\boldsymbol q}^\perp$ satisfy
\begin{equation}
    \label{eq:qde-information-duality-exp}
    \chi(W_\qv)=H(\qv),
    \qquad I(W_\qv^\perp) =\log p-H(\qv),
    \qquad \chi(W_\qv) + I(W_\qv^\perp) = \log p.
\end{equation}
Indeed, the outputs of $W_{\boldsymbol q}$ are pure and their average is $\frac1p\sum_d\ket{\phi_d(\boldsymbol q)}\bra{\phi_d(\boldsymbol q)}=\mathrm{diag}(\boldsymbol q)$, while $W_{\boldsymbol q}^\perp$ adds to a uniform input independent noise of entropy $H(\boldsymbol q)$.
Here $\chi(W_{\boldsymbol q})$ is the Holevo information of the BPQM message, which bounds how much information about the code symbol $d$ can be extracted from $\ket{\phi_d(\boldsymbol q)}$. It attains its maximum $\log p$ only at the uniform distribution, $\boldsymbol q=\boldsymbol u$, where the states are orthogonal and $P_{\mathrm{err}}(\boldsymbol q)=0$. 
On the other side, $I(W_{\boldsymbol q}^\perp)$ is the information that the output $Y=X+A$ of the dual classical channel carries about its input $X$.
The proof below identifies this classical dual $W_\qv^\perp \equiv \mathcal{L}(\qv)$ with the Gibbs reconstruction posterior law in \cref{sec:reconstruction}.
Since $\chi+I=\log p$, the quantum message is fully informative exactly when the dual classical observation is uninformative: BPQM decodes the root with vanishing error precisely when tree reconstruction is impossible.

\begin{proof}[Proof of \cref{thm:main}]
To prove \cref{eq:main-finite-depth}, we establish an algebraic equivalence between the BPQM message recursion defined in \cref{eq:qde-symbol,eq:qde-outcomes,eq:qde-check} of \cref{sec:qde} and the reconstruction posterior distributional recursion defined in \cref{eq:app-linsat-de,eq:reconstruction-convolution} of \cref{sec:reconstruction}.

We prove the first identity in \cref{eq:main-finite-depth} by induction, using the identities \eqref{eq:qde-dual-convolution} and \eqref{eq:qde-dual-product} established above. Suppose the inductive hypothesis $\mathcal L(\boldsymbol q^{(\ell)}) \overset{d}{=} \widetilde{\boldsymbol\eta}^{(\ell)}$ at depth $\ell$. At $\ell=0$ both sides equal $\boldsymbol e_0$, since $\mathcal L(\boldsymbol e_0)=\boldsymbol e_0$. Assume $\mathcal L(\boldsymbol q^{(\ell)})\overset{d}{=}\widetilde{\boldsymbol\eta}^{(\ell)}$. One BPQM round (\cref{fig:qde-round}) is an equality update at each child symbol followed by a parity update at the parent check. A BPQM symbol is a max-LINSAT constraint $j_t$ and a BPQM check is a max-LINSAT variable $i$, so these two updates should become the constraint stage and the variable stage of the reconstruction recursion, respectively. We verify this by lifting each update.

\emph{Equality update at symbol $t$.} By \cref{eq:qde-symbol}, $\boldsymbol v_t^{(\ell)}=\boldsymbol\mu*\boldsymbol q_{t,1}^{(\ell)}*\cdots*\boldsymbol q_{t,k-1}^{(\ell)}$, where the $\boldsymbol q_{t,h}^{(\ell)}$ are independent copies of $\boldsymbol q^{(\ell)}$ coming from disjoint subtrees. By \cref{eq:qde-dual-convolution}, $\mathcal L(\boldsymbol v_t^{(\ell)})$ is the convolution of the independent lifts of $\boldsymbol\mu$ and of the $\boldsymbol q_{t,h}^{(\ell)}$. By the inductive hypothesis, the convolution of the last $k-1$ lifts has the law of $\boldsymbol G_t$ in \cref{eq:reconstruction-convolution}, while $\mathcal L(\boldsymbol\mu)_x=\mu_{U_t-x}$ with $U_t\sim\boldsymbol\mu$. Using the convolution convention of \cref{eq:bpqm-equality} and $\mu_a=w_\beta(a)/c_\beta$ from \cref{eq:reconstruction-weight},
\begin{equation}
    \label{eq:main-factor-lift-exp}
    \mathcal L(\boldsymbol v_t^{(\ell)})_x
       \overset{d}{=}\sum_{w\in\mathbb F_p}G_t(w)\,\mu_{U_t-x+w}
       =\frac{e^{-\beta}+(1-e^{-\beta})\,\Gamma_t(U_t-x)}{c_\beta},
\end{equation}
where the last step uses $\sum_wG_t(w)\mathbf 1\{s+w\in S\}=\Gamma_t(s)$. Up to the constant $1/c_\beta$, this is the likelihood contributed by constraint $j_t$ in the reconstruction recursion.

\emph{Parity update at the parent check.} The parent check is a variable $i$ with parent constraint $j_0$ and child constraints $j_1,\ldots,j_{D-1}$. Its parity condition $\sum_jB_{ji}d_j=0$ reads $d_{j_0}=\sum_t\alpha_td_{j_t}$ with $\alpha_t=-B_{j_ti}/B_{j_0i}=-\gamma_t$. \Cref{eq:qde-outcomes,eq:qde-check} relabel each input as $\boldsymbol\nu_t=\mathcal R_{\alpha_t}\boldsymbol v_t^{(\ell)}$ and then apply the parity rule $D-2$ times. By the relabeling identity after \cref{eq:qde-dual-product}, $\mathcal L(\boldsymbol\nu_t)_z\overset{d}{=}\mathcal L(\boldsymbol v_t^{(\ell)})_{-\gamma_tz}$, and by \cref{eq:qde-dual-product} the parity updates lift to a normalized product:
\[
    \mathcal L(\boldsymbol q^{(\ell+1)})_z
      \overset{d}{=}
      \frac{\prod_{t=1}^{D-1}\mathcal L(\boldsymbol v_t^{(\ell)})_{-\gamma_tz}} {\sum_{y\in\mathbb F_p}\prod_{t=1}^{D-1}\mathcal L(\boldsymbol v_t^{(\ell)})_{-\gamma_ty}}.
\]
Substituting \cref{eq:main-factor-lift-exp} with $x=-\gamma_tz$, the constants $1/c_\beta$ cancel and each factor becomes $e^{-\beta}+(1-e^{-\beta})\Gamma_t(U_t+\gamma_tz)$. The $U_t$ and $\boldsymbol G_t$ are independent across $t$, and the $\gamma_t$ are the coefficient ratios drawn from $\Lambda$, exactly as in \cref{eq:app-linsat-de}. Hence $\mathcal L(\boldsymbol q^{(\ell+1)})\overset{d}{=}\widetilde{\boldsymbol\eta}^{(\ell+1)}$, which completes the induction.

To show the second identity in \cref{eq:main-finite-depth}, we apply \cref{eq:reconstruction-order} and the first identity, averaging over $A\sim\boldsymbol q^{(\ell)}$ given $\boldsymbol q^{(\ell)}$,
\[
    \Psi_\ell+\frac1p
      =\mathbb E\big[\mathcal L(\boldsymbol q^{(\ell)})_0\big]
      =\mathbb E\big[q^{(\ell)}_A\big]
      =\mathbb E\Big[\sum_a\big(q^{(\ell)}_a\big)^2\Big],
\]
and $\sum_a(q_a-1/p)^2=\sum_aq_a^2-1/p$ gives $\Psi_\ell=\mathbb E\big[\|\boldsymbol q^{(\ell)}-\boldsymbol u\|_2^2\big]$.

We now prove the threshold equality \eqref{eq:main-threshold-equality}. Write $\widehat q(b)=\sum_aq_a\omega^{ab}$ and
$m_\ell(b)=\mathbb E[|\widehat q^{(\ell)}(b)|^2]$.
Parseval's identity, the root convolution \cref{eq:qde-root}, and independence of its $k$ inputs give
\begin{align}
    \Psi_\ell &=\frac1p\sum_{b\ne0}m_\ell(b), \nonumber \\
    \mathbb E\!\left[
       \|\boldsymbol\mu_{\mathrm{root}}^{(\ell)}-\boldsymbol u\|_2^2\right]
       &=\frac1p\sum_{b\ne0}|\widehat\mu(b)|^2m_\ell(b)^k.
       \label{eq:main-root-fourier}
\end{align}
For $\beta>0$, each $\widehat\mu(b)$, $b\ne0$, is nonzero: it is a positive multiple of $\sum_{a\in S}\omega^{ab}$, which cannot vanish for prime $p$ and a nonempty proper $S$. Indeed, the minimal polynomial of $\omega^b$ is $1+X+\cdots+X^{p-1}$, whereas $\sum_{a\in S}X^a$ is not a multiple of it. The two nonnegative finite sums therefore vanish together as $\ell\to\infty$. The error function \cref{eq:bpqm-error} is continuous and vanishes only at $\boldsymbol u$, so the compactness of the probability simplex gives
\begin{equation}
    \label{eq:main-root-criterion}
    \lim_{\ell\to\infty} P_{\mathrm{err}}^{(\ell)}=0
       \quad\Longleftrightarrow\quad
    \Psi_\infty = 0.
\end{equation}
At $\beta=0$, $\boldsymbol\mu=\boldsymbol u$ and both conditions hold directly. This proves the threshold equality \eqref{eq:main-threshold-equality}.

Finally, for any fixed $\beta < \beta_{\mathrm{dyn}}$, we have $\Psi_\infty=0$, which by the second identity in \cref{eq:main-finite-depth} is exactly the hypothesis $\mathbb E\|\boldsymbol q^{(\ell)}-\boldsymbol u\|_2^2\to0$ of \cref{cor:finite-sampling}. That corollary yields the polynomial-time coherent Gibbs sampling statement.
\end{proof}

\begin{corollary}[Information monotonicity in inverse temperature]
\label{cor:beta-monotonicity}
Fix $(k,D,p,S,\Lambda)$ as in \cref{sec:ensemble}. For every depth $\ell\geq0$, the reconstruction information $I_\ell(\beta)$ in \cref{eq:reconstruction-definition} is nondecreasing in $\beta\geq0$. The same holds for its limit as $\ell\to\infty$.
\end{corollary}

\begin{proof}
Write $\boldsymbol s$ for the uniform distribution on $S$, and $\boldsymbol u$ for the uniform distribution on $\mathbb F_p$. The physical channel vector in \cref{eq:reconstruction-weight} satisfies
\[
 \boldsymbol\mu_\beta=(1-t_\beta)\boldsymbol u+t_\beta\boldsymbol s,
 \qquad
 t_\beta=\frac{r(1-e^{-\beta})}{r+(p-r)e^{-\beta}},
 \quad r=|S|.
\]
Since $t_\beta$ increases with $\beta$, for $0\leq\beta_1<\beta_2$ we may set $a=t_{\beta_1}/t_{\beta_2}\in[0,1]$ and $\boldsymbol\nu=a\boldsymbol e_0+(1-a)\boldsymbol u$. Convolution with $\boldsymbol u$ gives $\boldsymbol u$, so
\[
 \boldsymbol\mu_{\beta_1}=\boldsymbol\mu_{\beta_2}*\boldsymbol\nu.
\]
The equality rule \cref{eq:bpqm-equality-unitary} therefore gives
\[
 V_{\rm eq}
 \bigl(\ket{\phi_d(\boldsymbol\mu_{\beta_2})}
       \otimes\ket{\phi_d(\boldsymbol\nu)}\bigr)
 =\ket{\phi_d(\boldsymbol\mu_{\beta_1})}\otimes\ket0.
\]
Thus, given the physical output at $\beta_1$, we can append $\ket0$, apply $V_{\rm eq}^{\dagger}$, and discard the second system to obtain exactly the output at $\beta_2$, without knowing the codeword symbol $d$.

Consider the quantum combining tree dual to the depth-$\ell$ reconstruction observation, and let $\chi_\ell(\beta)$ denote its Holevo information, including the heralded labels. 
Apply the above operation to each physical output, keeping the code constraints fixed. This converts the entire quantum observation at $\beta_1$ into the observation at $\beta_2$. The BPQM representation retains all information about its input: equality combining is a unitary compression, while at a parity node averaging the child symbols with fixed sum removes coherences between distinct herald outcomes, so retaining the herald label loses no information. Applying the data processing inequality therefore gives $\chi_\ell(\beta_1)\geq\chi_\ell(\beta_2)$.

By the finite-depth posterior correspondence proved above and \cref{eq:qde-information-duality-exp}, averaging over the herald labels and known tree data gives
\[
 I_\ell(\beta)=\log p-\chi_\ell(\beta).
\]
Hence $I_\ell(\beta_1)\leq I_\ell(\beta_2)$. Taking $\ell\to\infty$ proves the limiting statement.
\end{proof}

\paragraph{Example: Binary case.}
For $p=2$ with $S=\{0\}$, the physical channel and its overlap are
\begin{equation}
    \label{eq:qde-binary-temperature}
    \boldsymbol\mu_j=\frac{(1,e^{-\beta})}{1+e^{-\beta}},
    \qquad \eta=\mu_{j,0}-\mu_{j,1}=\tanh(\beta/2).
\end{equation}
In this case the posterior magnetization, meaning its probability of $0$ minus its probability of $1$, is
\begin{equation}
    \label{eq:qde-binary-lift}
    M=\mathcal L(\boldsymbol q)_0-\mathcal L(\boldsymbol q)_1
      =(-1)^A(q_0-q_1),
    \qquad \mathbb E[M\mid\boldsymbol q]=(q_0-q_1)^2.
\end{equation}
Averaging this magnetization is therefore equivalent to tracking the mean-square quantum overlap. The factor of two in \cref{eq:qde-binary-temperature} follows from our convention that $\beta$ multiplies the number of violated constraints; the spin-coupling convention uses $\beta/2$.

By \cref{eq:qde-binary-lift,eq:main-finite-depth}, $2\Psi_\ell=\mathbb E[(q_0^{(\ell)}-q_1^{(\ell)})^2]$. With $\eta=\tanh(\beta/2)$, \cref{eq:main-root-fourier} and the binary specialization of \cref{eq:bpqm-error} yield
\begin{equation}
    \label{eq:main-binary-error}
    \frac{\eta^2}{4}(2\Psi_\ell)^k
       \leq P_{\mathrm{err}}^{(\ell)}
       \leq \frac{\eta^2}{2}(2\Psi_\ell)^k.
\end{equation}
To obtain these bounds, use $u/2\leq1-\sqrt{1-u}\leq u$ for $0\leq u\leq1$ on the squared root overlap.

\section{Computational Experiments}
\label{sec:experiments}

\newcommand{\SectionElevenGridRows}{525}
\newcommand{\SectionElevenFiniteRows}{252}
\newcommand{\SAHighMedianAbsError}{0.273}
\newcommand{\SAHighMeanAbsError}{0.582}
\newcommand{\SAHighMaxAbsError}{2.815}
\newcommand{\SAHighWithinTwo}{233}
\newcommand{\SALowMedianAbsError}{0.873}
\newcommand{\SALowMeanAbsError}{0.920}
\newcommand{\SALowMaxAbsError}{2.158}
\newcommand{\SALowWithinTwo}{249}

In the preceding sections we have obtained a rigorous proof that Regev+BPQM efficiently produces coherent Gibbs samples for every $\beta < \beta_\dyn$ for average-case $D$-regular max-$k$-LINSAT.
It has been conjectured \cite{KT87dynamics,MS06,Alaoui2025sampling,ZSSJ26} that classical Markov chain methods using single-symbol updates to sample the Gibbs distribution encounter slow mixing at inverse temperatures above the reconstruction threshold $\beta_{\mathrm{dyn}}$. This conjecture suggests that widely-used optimization methods such as simulated annealing and Glauber dynamics should fail efficiently reach temperatures below $1/\beta_{\mathrm{dyn}}$. In this section, we perform some computer experiments to test this conjecture. Everything reported in the preceding sections, showing the equivalence between the lowest temperature reachable by Regev+BPQM  and $\beta_{\mathrm{dyn}}$ was mathematically rigorous. Everything in this section, regarding the equivalence between the lowest energy efficiently reachable by simulated annealing and the expected energy at $\beta_{\mathrm{dyn}}$ is purely empirical. 

Note that $\beta_{\mathrm{dyn}}$ should not be regarded as a computational hardness threshold in all parameter regimes. In \cite{ZSSJ26} it was shown that for sufficiently large $k$ and $D$, a variant of Prange's algorithm can sample at temperatures below $1/\beta_{\mathrm{dyn}}$ in polynomial time. Nevertheless, at small $k$ and $D$, simulated annealing is difficult to beat in practice and we are not aware of any counterexample disproving the notion that $\beta_{\mathrm{dyn}}$ does represent the classical computational hardness barrier at sufficiently small $k$ and $D$.

In what follows, we report on computational experiments for $D$-regular max-$k$-LINSAT over an alphabet of size $p$ with target sets of size $r$. Thus, we explore instances across a four-dimensional parameter space. For a given choice of $(k,D,p,r)$ we can compute the corresponding value of $\beta_{\mathrm{dyn}}$ and the resulting expected fraction of constraints satisfied in the Gibbs distribution. We then compare this against the satisfaction fraction achieved by simulated annealing with a large number of sweeps and optimized hyperparameters. Remarkably, we find that throughout the entire four-dimensional parameter region we study, the discrepancy between these two numbers is on average only $0.58$ percentage points and at most $2.8$ percentage points.
Being subject to finite-size effects and statistical error bars, our computer experiments cannot be definitive. Nevertheless, we consider it unlikely that the consistently close agreement of these satisfaction fractions, as they trace out a complicated function on a four dimensional space, is coincidental. In \cite{piveteau2026scirate} and \cite{MPRP26}, it was previously observed that Regev+BPQM closely matches the performance of simulated annealing. However, the reason behind this remained a mystery prior to the present work.

\paragraph{Instance ensemble.}
For the computational experiments, take the common target set in \cref{sec:ensemble} to be $S_r=\{0,1,\ldots,r-1\}$, with independent uniform offsets $\tau_a$ and targets $F_a=\tau_a+S_r$. Thus row $a$ is satisfied when
\begin{equation}
    (B\mathbf{x}-\boldsymbol{\tau})_a\in S_r.
    \label{eq:experiments-residual}
\end{equation}
The Tanner graph is sampled uniformly from the simple $(k,D)$-regular ensemble.
For each variable, we choose its coefficients for the $D$ incident constraints according to the law $\Lambda_{\rm perm}$.

The parameter grid is
\begin{equation}
    p\in\{2,3,5,7,11,13\},\qquad
    3\leq k<D\leq 8,\qquad
    1\leq r<p,
    \label{eq:experiments-grid}
\end{equation}
which contains $\SectionElevenGridRows$ settings. All finite-graph calculations use $n=6720$ optimization variables. For an assignment $\mathbf{x}$, write
\begin{equation}
    e(\mathbf{x})=\frac{E(\mathbf{x})}{m}
    \label{eq:experiments-energy}
\end{equation}
for its fraction of violated constraints. We restrict the comparison to the $\SectionElevenFiniteRows$ settings for which the density-evolution calculation detected a finite reconstruction threshold.
Note that on the tree, the residual for the $j$-th constraint $U_j =(B\xv-\boldsymbol{\tau})_j$ follows the distribution $\boldsymbol{\mu}$ in \eqref{eq:reconstruction-weight}. At $\beta=\beta_\dyn$, the expected fraction of violated constraints on the tree is thus
\begin{equation}
    e_{\rm dyn} 
       =\frac{(p-r)e^{-\beta_\dyn}}
                    {r+(p-r)e^{-\beta_\dyn}},
    \label{eq:experiments-edyn}
\end{equation}
which is also the equilibrium energy of finite random instances at any temperature in the replica-symmetric phase.

\paragraph{Numerical method for determining $\beta_\dyn$.}
We use the operational definition $\beta_\dyn$ in \eqref{eq:linsat-betadyn-explicit} in terms of the reconstruction distributional recursion. In practice, we determine $\beta_\dyn(k,D,q,S_r,\Lambda_{\rm perm})$ via the following planted density evolution algorithm:

\begin{enumerate}
    \item Initialize a pool of $M$ samples at the planted value, \ie,\ set $\{\tilde\etav^{(0)}_\mu = \ev_0 : \mu=1,2,\ldots,M\}$. By default, take $M=10^4$.
    \item For each step of the recursion \eqref{eq:app-linsat-de}, replace the pool by $M$ fresh samples as follows: For each sample, draw one $D$-tuple from $\Lambda_{\rm perm}$ and form all $D-1$ child-to-parent ratios $\gamma_t$; 
	then per constraint draw $k-1$ children uniformly from the current pool and a noise $U_t$. Convolve the children by multiplying length-$p$ FFTs, and accumulate the logarithm of the bracket over the $D-1$ constraints before exponentiating and normalizing. One sweep costs $O(MDk\,p\log p)$.
    \item Track the order parameter $\Psi_\ell$. Declare the trivial fixed point reached when it drops below $0.01\,(1-1/p)$, and a nontrivial one when the local slope $s_j = -\log(\Psi_{\ell_j}/\Psi_{\ell_{j-1}})/\log(\ell_j/\ell_{j-1})$, measured across geometric checkpoints $\ell_j\approx24\cdot2^{j/2}$, falls below $0.05$ twice in a row while $\Psi_\ell>0.05\,(1-1/p)$.
	Locate the threshold by bisection in $\beta$.
\end{enumerate}

\paragraph{Annealing experiments and comparison with reconstruction prediction.}
We ran two sets of simulated annealing experiments, with different sweep counts.
The high-sweep-count simulated-annealing calculation uses 32 graph instances per setting, one restart, $14.4\times10^6$ sweeps with a calibrated linear-$\beta$ schedule, and greedy coordinate updates to a one-coordinate local optimum. The low-sweep-count calculation uses $144{,}000$ sweeps and reports the best state encountered. These two budgets differ by a factor of $100$.

\begin{figure}[ht]
    \centering
    \begin{subfigure}[t]{0.49\linewidth}
        \centering
        \includegraphics[width=\linewidth]{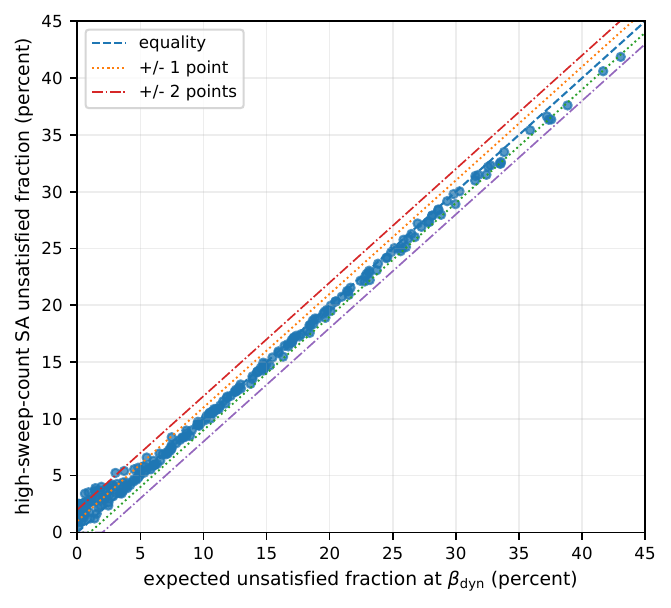}
        \caption{$14.4\times10^6$ sweeps.}
        \label{fig:sa-high-vs-edyn}
    \end{subfigure}\hfill
    \begin{subfigure}[t]{0.49\linewidth}
        \centering
        \includegraphics[width=\linewidth]{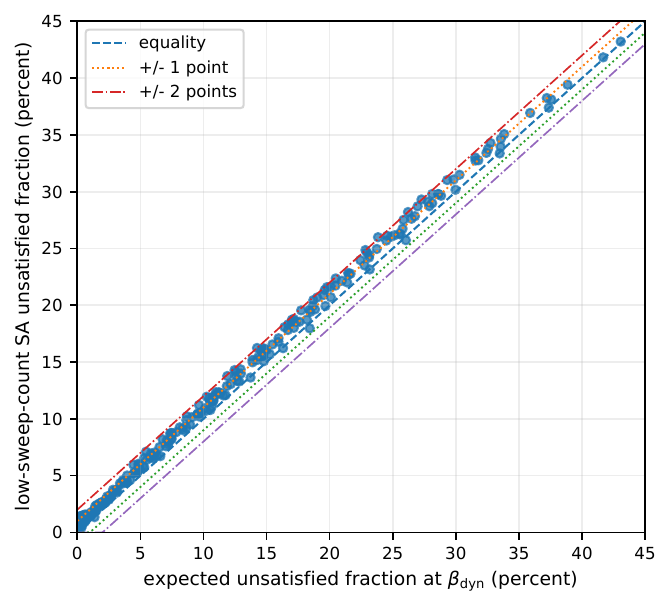}
        \caption{$144{,}000$ sweeps.}
        \label{fig:sa-low-vs-edyn}
    \end{subfigure}
    \caption{Simulated-annealing violation fractions versus the tree-based reference $e_{\rm dyn}$ in \cref{eq:experiments-edyn}. (a) Violation fraction reached by the high-sweep-count protocol. (b) Best violation fraction found by the low-sweep-count protocol. Both panels show the same $252$ parameter settings with detected finite reconstruction thresholds; in panel (a), each point is a mean over $32$ graph instances. The dashed line is equality; dotted and dash-dotted lines mark one- and two-percentage-point discrepancies. Reducing the sweep count by a factor of $100$ shifts the typical result upward by about one percentage point but preserves its dependence on the reconstruction prediction.}
    \label{fig:sa-vs-edyn}
\end{figure}

In both simulated-annealing protocols, the reported violation fraction closely follows $e_{\rm dyn}$ across the comparison grid; see \cref{fig:sa-vs-edyn}. The absolute discrepancies are summarized in \cref{tab:experiments-accuracy}. Every setting in both calculations is within three percentage points of $e_{\rm dyn}$.

The agreement is more specific than a high correlation between two nearly monotone quantities. The discrepancies remain small in absolute units across the four-dimensional grid $(k,D,p,r)$. The high-sweep-count calculation is within two percentage points at $\SAHighWithinTwo$ of the $\SectionElevenFiniteRows$ settings; the low-sweep-count calculation is within two points at $\SALowWithinTwo$ settings.

\begin{table}[ht]
    \centering
    \small
    \begin{tabular}{lrrrr}
        \hline
        annealing protocol & settings & median & mean & maximum \\
        \hline
        high-sweep-count SA & $\SectionElevenFiniteRows$ & $\SAHighMedianAbsError$ & $\SAHighMeanAbsError$ & $\SAHighMaxAbsError$ \\
        low-sweep-count SA & $\SectionElevenFiniteRows$ & $\SALowMedianAbsError$ & $\SALowMeanAbsError$ & $\SALowMaxAbsError$ \\
        \hline
    \end{tabular}
    \caption{Absolute discrepancy between the reported simulated-annealing violation fraction and the tree-based reference $e_{\rm dyn}$, in percentage points of violated constraints. The two protocols use $14.4\times10^6$ and $144{,}000$ sweeps, respectively.}
    \label{tab:experiments-accuracy}
\end{table}

\paragraph{Interpretation.}
Across the settings with a detected finite threshold, the reconstruction-based energy scale closely predicts the energies reached by simulated annealing at both tested budgets. This empirical agreement complements the tree-threshold identity in \cref{thm:main}. The finite-size optimization measurements do not establish convergence to the Gibbs distribution or a limit on what longer runs or other classical algorithms can achieve.

\FloatBarrier

\section*{Acknowledgments}
We thank Christophe Piveteau and Ruslan Shaydulin for sharing quantum density evolution data which provided some of the first hints from which this paper grew. We thank Mark Sellke and Noah Shutty for useful discussions. 
LZ is supported by Department of Navy Award No. N000142612399 issued by the Office of Naval Research.
\\

\noindent
\textbf{AI Disclosure:} We used Claude, Gemini, and ChatGPT to write code, analyze data, develop proofs from our sketches, and assist with drafting and editing the manuscript. All aspects of the project were overseen by the authors, and we take responsibility for the correctness of all claims.

\begingroup
\small
\bibliography{rrrr}
\endgroup

\end{document}